\documentclass[twocolumn]{revtex4}
\usepackage{graphicx,epsf}
\usepackage{subfigure}
\usepackage{color}
\usepackage{dcolumn}
\usepackage{titlesec}
\usepackage{amsfonts}
\usepackage{mathrsfs}
\usepackage{amsmath}

\usepackage{subfig}

\begin{document}

\title{Kinetic theory of geodesic acoustic modes in toroidal plasmas: a brief review}

\author{Zhiyong Qiu$^{1}$, Liu Chen$^{1, 2}$ and Fulvio Zonca$^{3, 1}$}

\affiliation{$^1$Institute for    Fusion Theory and Simulation and Department of Physics, Zhejiang University, Hangzhou, P.R.C\\
$^2$Department of   Physics and Astronomy,  University of California, Irvine CA 92697-4575, U.S.A.\\
$^3$ ENEA, Fusion and Nuclear Safety Department,
C. R. Frascati, Via E. Fermi 45, 00044 Frascati (Roma), Italy}

\begin{abstract}
{Geodesic acoustic modes (GAM) are oscillating zonal structures unique to toroidal plasmas, and have been extensively studied in the past decades due to their potential capabilities of regulating microscopic turbulences and associated anomalous transport. This article reviews linear and nonlinear theories of GAM; with emphases on kinetic treatment, system nonuniformity and  realistic magnetic geometry, in order to reflect  the realistic  experimental conditions.  Specifically, in the linear physics, the resonant wave-particle interactions are discussed, with the application to resonant excitation by energetic particles (EPs). The theory of  EP-induced GAM (EGAM) is applied to realistic devices for the interpretation of   experimental observations, and global effects due to coupling to GAM continuum are also discussed.      Meanwhile, in the nonlinear physics, the spontaneous GAM  excitation by microscale turbulences is reviewed, including the effects of various system nonuniformities. A unified theoretical framework of GAM/EGAM is then constructed based on our present understandings.  The first-principle-based GAM/EGAM theories  reviewed here, thus, provide  the tools needed for the understanding and interpretation of experimental/numerical results. }
\end{abstract}

\maketitle


\section{Introduction}

The peculiar role  of toroidally and poloidally symmetric zonal structures  (ZS) \cite{AHasegawaPoF1979,MRosenbluthPRL1998,DSpongPoP1994,FZoncaNJP2015,PDiamondPPCF2005} and their influence on the overall plasma performance has been well accepted and extensively studied in the past two decades.  ZS can regulate  microscale drift wave turbulence (DW) \cite{WHortonRMP1999}, including drift Alfv\'en waves (DAWs),  via scattering into short radial wavelength stable domain, and thereby, suppress the DW induced anomalous transport.  ZS can, thus, be viewed as the generator of nonlinear equilibria  with suppressed turbulence \cite{ADimitsPoP2000,LChenNF2007}, and possibly an important factor in the H-mode confinement \cite{FWagnerPRL1982,FWagnerPPCF2007}.

Geodesic acoustic modes (GAM)  \cite{NWinsorPoF1968,FZoncaEPL2008}, as the finite frequency counterpart  of zonal flow, have been observed  in various machines by different diagnostics  \cite{MJakubowskiPRL2002,AFujisawaPRL2004,AMelnikovPPCF2006,TIdoNF2006,GConwayPPCF2005,ALiuPRL2009,ALiuPPCF2010,KZhaoPRL2006,DKongNF2013,DKongNF2013b} in the search of zero frequency zonal flow (ZFZF) \cite{MRosenbluthPRL1998}, with the linear features such as mode  frequency,  three dimension mode structure, density perturbation and radial propagation identified.  An inverse relation  of turbulence level and GAM intensity were often observed,  suggesting the GAMs are excited nonlinearly by ambient turbulence, as  shown by   bicoherence analysis \cite{GXuPRL2003,YNagashimaPRL2005,TLanPPCF2008}.  Theoretically, the regulation of  DW by ZFZF and/or GAM, is achieved via the spontaneous excitation of ZFZF/GAM by DWs  modulational instability \cite{LChenPoP2000,FZoncaEPL2008}. Thus, the nonlinear drive from DWs  in the form of   Reynolds stress \cite{YXuPRL2000} must be  strong enough to overcome the threshold conditions due to frequency mismatch and/or dissipations. The nonlinearly generated ZFZF/GAM,  in turn, scatter DWs into stable short wavelength domain.    Noting the  fact that  both ZFZF and GAM can be excited by and regulate DWs, and that their respective  nonlinear coupling   cross-sections  based on gyrokinetic predictions  are comparable \cite{LChenPoP2000,FZoncaEPL2008},  understanding the nonlinear dynamics of DWs and, thus, quantitative prediction of the transport level  require  careful examination of linear drive/dissipations of GAM, and the possible direct power transfer between GAM and ZFZF.

Due to its finite frequency, GAM can resonate with, and be excited by energetic particles (EPs) \cite{HBerkNF2006,RNazikianPRL2008}, analogous to the shear Alfv\'en wave (SAW) continuum mode excitation by EPs \cite{LChenPoP1994}.  Though EGAM typically has a radial scale much longer than that of GAM driven by DWs, the possible nonlinear interactions between EGAM and DWs \cite{DZarzosoPRL2013,RDumontPPCF2013} were observed numerically, suggesting EGAM as an active control for DWs. The observed oscillations at twice of GAM/EGAM frequency \cite{RNazikianPrivate2009}, furthermore, suggest the nonlinear self-couplings of GAMs, including generating GAM/EGAM second harmonic and ZFZF,  as demonstrated by numerical simulations \cite{HZhangNF2009}. The generation of GAM/EGAM second harmonic \cite{HZhangNF2009,GFuJPP2011,ZQiuPoP2017}, as an additional dissipation mechanism for GAM/EGAM, and generation of ZFZF as a channel for direct power transfer from GAM/EGAM to ZFZF \cite{LChenEPL2014,ZQiuPoP2017}, will affect the branching ratio of GAM and ZFZF generation by DWs, and, as a consequence,  DWs nonlinear dynamics.

In this paper, the theoretical investigation of GAM is briefly reviewed, with emphasis  on first-principle-based kinetic treatment and realistic geometry. Therefore,  the present result can be directly applied to explain experimental observations and numerical simulations in the the proper limits.  The rest of the paper is organized as follows. In Sec. \ref{sec:GAM_linear}, the linear properties of GAM are presented, with the fluid derivation  and the discussions of GAM continuum given in \ref{sec:GAM_fluid}, and the kinetic treatment given in Sec. \ref{sec:GAM_kinetic_theory},  emphasizing on the physics picture of wave-particle resonances in the short wavelength limit. The EGAM local and global theories are reviewed in Sec. \ref{sec:EGAM}, with   applications to several specific cases in realistic  devices .  Speculations are made on EGAM nonlinear saturation and possible particle losses  due to EGAM induced pitch angle scattering. The spontaneous excitation of GAM by DWs is reviewd in Sec. \ref{sec:NL_GAM_DW}, taking ion temperature gradient  (ITG) DW in the local limit as an example. The local nonlinear  theory is then generalized to GAM excitation by short wavelength collisionless trapped electron mode (CTEM) DW and toroidal Alfv\'en eigenmode (TAE). Further extension  to global theory is also given,  considering the system nonuniformities. The GAM/EGAM nonlinear self-coupling, as one important factor for the nonlinear DW dynamics, is also reviewed in Sec. \ref{sec:self_NLty}.  In Sec. \ref{sec:framework}, a unified theoretical framework of GAM/EGAM is proposed, including the main processes discussed in this paper.   Conclusions and discussions are given in Sec. \ref{sec:conclusion}.

\section{Linear theory of GAM}
\label{sec:GAM_linear}

In this section, we present the linear GAM theory. First, in Sec. \ref{sec:GAM_fluid}, adopting a fluid approach to illustrate fundamental properties of the GAM continuous spectrum. Afterwards, in Sec. \ref{sec:GAM_kinetic_theory}, introducing kinetic description and the properties of GAM at short wavelengths.

\subsection{Fluid theory:  GAM continuum and mode conversion to kinetic GAM}
\label{sec:GAM_fluid}

In Sec. \ref{sec:GAM_fluid}, the fluid theory of GAM will be presented \cite{NWinsorPoF1968,ZQiuPST2011}, with the  GAM continuum due to plasma nonuniformity \cite{FZoncaEPL2008,ZQiuPST2011} briefly reviewed,  as a peculiar feature of GAM.  The GAM continuum induced linear absorption \cite{FZoncaEPL2008} and the multiple-scale radial structure \cite{ZQiuPST2011} have important consequences on the  linear decay due to both continuum   and Landau damping \cite{ZQiuPPCF2009,ZQiuPST2011,FPalermoEPL2016,ABiancalaniPoP2016},  resonant excitation by EPs \cite{FZoncaIAEA2008,ZQiuPPCF2010,ZQiuPoP2012,LChenPoP2017}  and nonlinear interactions with DW/DAWs \cite{ZQiuPoP2014}, as we will discuss in the rest of the paper.  A thorough  and detailed investigation  of GAM continuous spectrum, including phase mixing and mode conversion to kinetic GAM (KGAM), was presented in Ref. \cite{ZQiuPST2011}.

We start with the linearized fluid equations,
\begin{eqnarray}
&&\partial_t\delta n+\nabla\cdot(n_0\delta\mathbf{v})=0\label{eq:fluid_continuity},\\
&&m_in_0\partial_t\delta\mathbf{v}=-\nabla\delta P+\delta\mathbf{J}\times \mathbf{B}_0/c\label{eq:fluid_momentum},\\
&&\delta P=\Gamma_eT_e\delta n_e+\Gamma_iT_i\delta n_i\label{eq:fluid_state},\\
&&\delta 
\mathbf{E}+\delta\mathbf{v}\times\mathbf{B}_0/c=0\label{eq:ohmlaw},
\end{eqnarray}
where  equations   (\ref{eq:fluid_continuity}), (\ref{eq:fluid_momentum}),  (\ref{eq:fluid_state}) and  (\ref{eq:ohmlaw}) are, respectively, linearized   continuity equation,  momentum equation, equation of state and Ohm's law;  $\delta n$  is the number density,  $\delta\mathbf{v}$ is perturbed velocity, $\Gamma$ is the appropriate ratio of specific heats, $T$ is the temperature; subscripts $e, i$ denote, respectively, electron and ion species. Other notations are standard. 

The governing GAM equation is derived from  the flux surface averaged quasineutrality condition,
\begin{eqnarray}
\partial_r\overline{\delta J_r}=0,\label{eq:qn_fluid}
\end{eqnarray}
with $\overline{(\cdots)}\equiv \int^{2\pi}_0 (\cdots)d\theta/2\pi$ denoting  magnetic surface averaging and  the perturbed radial current $\delta J_r$ obtained  from the poloidal component of  momentum  equation  as
\begin{eqnarray}
\delta J_r=(c/B_0)\left[n_0m_i\partial_t\delta v_{\theta}+(1/r)\partial_{\theta}\delta P\right].\label{eq:deltaJ_r}
\end{eqnarray}
Note that in Eq. (\ref{eq:qn_fluid}) we have neglected equilibrium nonuniformity scale with respect to GAM wavelength by dropping the Jacobian of the adopted toroidal flux coordinates that we use throughout this work. Equation (\ref{eq:deltaJ_r}) consists of two terms,  corresponding to,  respectively,  the  polarization current due to finite GAM frequency, and the   perturbed diamagnetic current associated with the perturbed pressure gradient in poloidal direction. $\delta v_{\theta}$ is the     GAM  radial electric field   induced poloidal  drift velocity  (``zonal flow"),  and the perturbed pressure $\delta P$ is obtained from equation of state, with the perturbed density   $\delta n$  given  by  the plasma compressibility due to toroidicity, noting the GAM radial wavelength is much shorter than equilibrium scale
\begin{equation}
\delta n=-\frac{cn_0k_G\overline{\delta\phi}_G\sin\theta}{\omega
B_0R_0}.\label{eq:density_perturbation}
\end{equation}
Equation (\ref{eq:density_perturbation}) is the well-known ``upper-down anti-symmetric" density perturbation of GAM in the fluid limit \cite{HZhaoPST2010}. Combining equations (\ref{eq:fluid_state}), (\ref{eq:qn_fluid}), (\ref{eq:deltaJ_r}) and (\ref{eq:density_perturbation}),  the radial GAM mode equation   can be derived  as:
\begin{equation}
\frac{\partial}{\partial
r}\left[\frac{c^2}{B^2_0}m_in_0\omega\left(1-\frac{\omega^2_G}{\omega^2}\right)\right]\frac{\partial}{\partial
r}\overline{\delta\phi}_G=0,\label{eq:GAM_eq_lowest_order}
\end{equation}
with $\omega^2_{G}\equiv(\Gamma_iT_i+\Gamma_eT_e)/(m_iR^2_0)$ being the GAM frequency in the fluid limit. Note that the coefficient of the highest order derivative can vanish, and thus, the equation is singular at $r_0$ with $\omega^2_{G}(r_0)=\omega^2$, suggesting the existence of GAM continuum \cite{FZoncaEPL2008}, similar to the well-known  shear Alfv\'en resonance \cite{LChenPoF1974,AHasegawaPoF1976}.  

Equation (\ref{eq:GAM_eq_lowest_order}) can be solved and yield the following solution,
\begin{eqnarray}
\delta E_G&=&A_+\exp(i\omega_G(r)t)+A_-\exp(-i\omega_G(r)t)\nonumber\\
&&+\frac{S_0\exp(-i\omega_0 t)}{\omega^2_0-\omega^2_G(r)},\label{eq:GAM_antenna_general}
\end{eqnarray}
in which   the homogeneous solutions correspond  to the initial perturbations of  GAM continuum, with $A_+$ and $A_-$ determined from initial condition, and the inhomogeneous solution  corresponds to   an  incoming oscillation, due to, e.g., an external antenna \cite{SItohPPCF2007,ZQiuPST2011}. This term also accounts for  EGAM driven away from $r_0$ \cite{FZoncaIAEA2008,ZQiuPoP2012} and/or nonlinear drive by DW/DAWs \cite{LChenVarenna2010} in the form of  ``$S_0\exp(-i\omega_0t)$".   Note that  the initial perturbation oscillates at the local GAM frequency $\omega_G(r)$ , and two nearby points initially with the same phase  will develop a phase difference $\omega'_G(r)\Delta r t$ in time, with $\Delta r$ being the radial distance and $\omega'_G\equiv\partial_r\omega_G$. Consequently, the radial wavenumber $\sim \omega'_G(r) t$ increases with time, and generates singular mode structures asymptoticly, leading to the phase mixing of  $\delta\phi_G\propto 1/t$ \cite{LChenPoF1974}.  On the other hand,  the  oscillation energy piles up at $r_0$, with the mode structure proportional to $1/(r-r_0)$ near the resonant point $r_0$. Finite absorption of the driving mode energy density by the plasma then  occurs and is described by the Poynting flux into the narrow singular layer around $r_0$, with the absorption power give by equation (18) of Ref. \cite{ZQiuPST2011}. 

The singular mode structure  given in equation (\ref{eq:GAM_antenna_general}) also indicates the breakdown of the MHD treatment at very short radial scales, and the necessity of kinetic treatment. Inclusion of the finite ion Larmor radius effects (FLR) \cite{AHasegawaPoF1976,ZQiuPST2011} will  remove the singularity and introduce   mode conversion of the singular continuum solution at $r_0$ to outward propagating KGAM at Airy scales, as discussed in the case of EGAM driven by a spatially broad EP beam \cite{ZQiuPoP2012} in Sec. \ref{sec:EGAM_global_strong}. Interested readers may refer to Ref. \citenum{ZQiuPST2011} for a more thorough and detailed discussion of interesting physics associated with GAM continuous spectrum.

\subsection{Kinetic  dispersion relation, and Landau damping in the short wavelength limit}
\label{sec:GAM_kinetic_theory}

The real frequency of GAM given by fluid theory is   not satisfactory for explaining experimental results,  due to uncertainties induced by the closure with the  equation of state for a collisionless plasma; although the   dependence on parameters are   qualitatively correct. 
Some key physics, e.g., wave-particle resonances, are missing in fluid model, which, however, play  important role in the GAM related physics such as collisionless Landau damping and excitation by EPs as  discussed in Sec. \ref{sec:EGAM}.  In this Section, we briefly summarize the main steps in deriving  the GAM linear dispersion relation adopting the gyrokinetic framework, while interested readers   may refer to a systematic  derivation with rigorous orderings  presented in Ref. \cite{ZQiuPPCF2009}.  The particle responses derived here, will also be applied in later sections for the nonlinear GAM interactions with microscopic turbulences.

The  perturbed particle distribution function $\delta f$ can be expressed as
\begin{eqnarray}
\delta f_s=(e_s/m_s)\partial_E F_0\delta\phi+\exp[i(m_sc)/(e_sB^2)\mathbf{k} \times\mathbf{B} \cdot\mathbf{v}]\delta H,\nonumber
\end{eqnarray}
and the nonadiabatic particle responses  $\delta H$,   can be derived from  the general gyrokinetic equation \cite{EFriemanPoF1982}:
\begin{eqnarray}
\left(-i\omega+v_{\parallel}\partial_l+i\omega_d\right)\delta H_k&=&-i\frac{e_s}{m_s}QF_0J_k\delta L_k\nonumber\\
&-&\Lambda^{\mathbf{k}}_{\mathbf{k'},\mathbf{k''}}J_{k'}\delta L_{k'}\delta H_{k''}
\label{eq:NLGKE}.
\end{eqnarray}
Here, $\omega_d=(v^2_{\perp}+2
v^2_{\parallel})/(2 \Omega R_0)\left(k_r\sin\theta+k_{\theta}\cos\theta\right)$ is the magnetic drift frequency for a circular cross section large aspect ratio tokamak,  $l$ is the length along the equilibrium magnetic field line, $QF_0\equiv(\omega\partial_E-\omega_*)F_0$, $E=(v^2_{\perp}+v^2_{\parallel})/2$, $\omega_*$ is the diamagnetic drift frequncy with  $\omega_*F_0\equiv \mathbf{k}\cdot\mathbf{b}\times\nabla F_0/\Omega$, $J_k\equiv J_0(k_{\perp}\rho_L)$ with $J_0$ being the Bessel function of zero-index accounting for FLR effects, $\rho_L \equiv mcv_{\perp}/(eB)$ is the Larmor radius,  $\delta L=\delta\phi-v_{\parallel}\delta A_{\parallel}/c$, $\Lambda^{\mathbf{k}}_{\mathbf{k'},\mathbf{k''}}\equiv(c/B_0)\sum_{\mathbf{k}=\mathbf{k'}+\mathbf{k''}} \mathbf{b}\cdot\mathbf{k''}\times\mathbf{k'}$; and other notations are standard. The second term on the right-hand side of equation (\ref{eq:NLGKE}) is the   convective nonlinearity, which will be used in Sec. \ref{sec:NL_GAM_DW} for the nonlinear interactions between GAM and DW/DAW turbulences. This is the general form of the gyrokinetic equation in Fourier space \cite{EFriemanPoF1982}, and its simplified versions in various limits are used in different sections of this paper for the specific problems of interest; e.g., electro-static limit for linear theory of GAM/EGAM and their nonlinear interactions with DW turbulence, and electro-magnetic limit for the nonlinear GAM excitation by TAE.  Note that in Sec. \ref{sec:self_NLty}, where self couplings of GAM/EGAM  are reviewed, an extended version of equation (\ref{eq:NLGKE}) including parallel nonlinearity is used, which is usually neglected because it is typically higher order in the gyrokinetic expansion parameter, and its effect correspondingly enters on a longer time scale compared with that of the convective nonlinearity \cite{ABrizardPoP1995,TSHahmPoF1988}.

In this Section, for   GAM with $n=0$ and predominantly electro-static perturbation, one has  $v_{\parallel}\partial_l=(v_{\parallel}/qR_0)\partial_{\theta}$,  $\omega_*=0$, $\delta L_G= \delta\phi_G$, and $\omega_d=\omega_{dr}\equiv k_r(v^2_{\perp}+2
v^2_{\parallel})\sin\theta/(2 \Omega R_0)\equiv\hat{\omega}_{dr}\sin\theta $ accounting for radial magnetic drift associated with geodesic curvature.  Equation (\ref{eq:NLGKE}) in the linear limit,  reduces to 
\begin{eqnarray}
\left(-i\omega+\omega_{tr}\partial_{\theta}+i\omega_{dr}\right)\delta H_G&=&i\omega\frac{e_s}{m_s}\partial_E F_0J_G\delta \phi_G,\nonumber
\end{eqnarray}
and for thermal plasmas with Maxwellian distribution function, $\partial_E F_0=-(m_s/T_s)F_0$. The GAM equation is derived from the quasi-neutrality condition
\begin{eqnarray}
\frac{n_0e^2}{T_i}\left(1+\frac{T_i}{T_e}\right)\delta\phi_k=\sum_{s=e,i}\left\langle q_s J_k\delta H_s\right\rangle_k, \label{eq:QN}
\end{eqnarray}
with $\langle\cdots\rangle$ denoting velocity space integration.

For GAM with typically $\omega\sim v_{ti}/R_0$, electron response to GAM can be derived, noting $|\omega_{tr,e}|\gg|\omega_G|$, and one has
\begin{eqnarray}
\delta H^L_{G,e}=\frac{e}{T_e}F_0\overline{\delta\phi}_G,
\end{eqnarray}
which cancels the electron adiabatic contribution in the perturbed distribution function, as expected. 
  
Decomposing the GAM scalar potential as
\begin{eqnarray}
\delta\phi_G=\sum_m\delta\phi_{G,m} e^{im\theta},\nonumber
\end{eqnarray}
with $\delta\phi_{G,m}$ obtained from equation (\ref{eq:deltaphi_tilde}), the perturbed ion response to GAM, can be derived as \cite{ZGaoPoP2006}
 \begin{eqnarray}
 \delta H_{G,i}&=&-\frac{e}{m_i}\omega J_G\partial_EF_0 \sum_p  \sum_m\sum_l \nonumber\\
 &&\times\frac{i^{(p-l)} J_l(\hat{\Lambda}_d)J_p(\hat{\Lambda}_d)e^{i(m+l+p)\theta}\delta\phi_{G,m}}{\omega-(l+m)\omega_{tr}}.\label{eq:GAM_response_general}
 \end{eqnarray}
Here, we have assumed well circulating ions with constant $v_{\parallel}$, $|\omega|\gg\omega_{b,i}$ with $\omega_{b,i}\sim \sqrt{\epsilon}\omega_{tr,i}$ being the trapped ion bounce frequency and $\epsilon\equiv r/R_0\ll1$   the inverse aspect ratio, $\hat{\Lambda}_d\equiv \hat{\omega}_{dr}/\omega_{tr}\equiv k_r\hat{\rho}_d$ with $\hat{\rho}_d\equiv \hat{v}_d/\omega_{tr}$  the drift orbit width and $\hat{v}_d\equiv (v^2_{\perp}+2
v^2_{\parallel})/(2 \Omega R_0)$, and the 
$e^{-i\hat{\Lambda}_d\cos\theta}=\sum_l (-i)^l J_l(\hat{\Lambda}_d)e^{il\theta}$ expansion is applied to derive equation (\ref{eq:GAM_response_general}). Note that  equation (\ref{eq:GAM_response_general}) is the general particle response to GAM, and  it can be used to obtain EP response in Sec. \ref{sec:EGAM}.

Different orderings can be taken for non-resonant and resonant ions to further simplify the general respoinse of equation (\ref{eq:GAM_response_general}).  For non-resonant bulk ions, with $v\sim v_{ti}\equiv \sqrt{2T_i/m_i}$, we have $|\omega_{tr,i}/\omega_G|\sim 1/q\ll1$ and $|\omega_d/\omega|\sim k_r\rho_{ti}\ll1$.  Here, $\rho_{ti}\equiv m_icv_{ti}/(eB_0)$. As a result,   the mode structure and  dispersion relation of GAM determined by  non-resonant thermal plasma response  can  be derived by substituting the ion response, equation (\ref{eq:GAM_response_general}), into quasi-neutrality condition, and applying the $\omega_{tr}\ll\omega$ and $\hat{\Lambda}_d\ll1$ limits. One then derives, the Hermitian part of GAM dispersion function
\begin{eqnarray}
D_{\tiny{R}}&=&\left[1-\left(\frac{7}{4}+\tau\right)\frac{v^2_{ti}}{\omega^2R^2_0}+\hat{b}\frac{v^2_{ti}}{\omega^2R^2_0}\left(\frac{31}{16}+\frac{9}{4}\tau+\tau^2\right)\right.\nonumber\\
&-&\frac{v^4_{ti}}{\omega^4R^4_0q^2}\left(\frac{23}{8}+2\tau+\frac{\tau^2}{2}\right)\nonumber\\
&-&\left.\hat{b}\frac{v^4_{ti}}{\omega^4R^4_0}\left(\frac{747}{32}+\frac{481}{32}\tau+\frac{35}{8}\tau^2+\frac{\tau^3}{2}\right)\right]\hat{b}\label{eq:GAM_kinetic_DR_real}.
\end{eqnarray}
with the subscript $R$ denoting real part, $\tau\equiv T_e/T_i$,  and $\hat{b}\equiv k^2_{\perp}\rho^2_{ti}/2$. Equation (\ref{eq:GAM_kinetic_DR_real}),   is derived based on the   $|k_r\rho_L|\ll1$ and $1/q^2\ll1$ expansion, which is usually satisfied in the parameter region  where GAM related physics are important.
The perturbed GAM scalar potential, can then be derived from quasi-neutrality condition as
\begin{eqnarray}
 \delta\phi_G&=&\overline{\delta\phi}_G\left\{1-\left[1-\hat{b}\left(\frac{3}{2}+\tau\right)\right]\tau\frac{\omega_{dt}}{\omega}\sin\theta\right.\nonumber\\
&-&\left[\frac{7}{4}+\tau-\hat{b}\left(\frac{13}{4}+\frac{19}{4}\tau+2\tau^2\right)\right]\tau\frac{\omega^2_{dt}}{2\omega^2}\cos2\theta\nonumber\\
&-&\left[\frac{9}{4}+\frac{7}{8}\tau-\left(\frac{9}{4}+\frac{7}{4}\tau+\frac{\tau^2}{2}\right)\cos2\theta\right]\tau\frac{\omega^3_{dt}}{\omega^3}\sin\theta\nonumber\\
&-&\left.\left(\frac{\tau^2}{2}+\tau\right)\frac{\omega_{dt}\omega^2_{tt}}{\omega^3}\sin\theta\right\},\label{eq:deltaphi_tilde}
\end{eqnarray}
with the terms proportional to $\hat{b}$ accounting for FLR effects, $\omega_{dt}/\omega$ for  FOW effects and $\omega_{tt}/\omega$ for  parallel ion compressibility. Here, $\omega_{dt}\equiv k_r\rho_{ti}v_{ti}/R_0$ and $\omega_{tt}\equiv v_{ti}/(qR_0)$.

The collisionless Laudau damping of the toroidally symmetric GAM,  is mainly induced by the thermal ion transit harmonic resonances. Noting the $\omega\gg\omega_{tr,i}\gg\omega_{b,i}$ ordering,  the ``number" of transit harmonics involved in the process is related to the ratio of GAM wavelength compared to the  ion  drift orbit width, as demonstrated by equation (\ref{eq:GAM_response_general}).   The  Landau damping of GAM due to primary transit resonance ($|\omega|=|\omega_{tr}|$), was investigated in Ref. \citenum{FHintonPPCF1999},  which was then extended to   small but finite drift orbit width regime, with $|\omega|=2|\omega_{tr}|$ resonances taken into account \cite{HSugamaJPP2006}. It was shown by TEMPEST simulations \cite{XXuPRL2008,XXuNF2009} that  higher order transit harmonic resonances becomes increasingly  more  important as one further increases $|k_r\hat{\rho}_d|$ (e.g., by increasing $q$ at fixed  $k_r\rho_L$ \cite{XXuPRL2008,XXuNF2009}). Therefore, it was noted that  the ``number" of particles that resonate  with $\omega=(l+m)\omega_{tr}$ transit harmonic is proportional to $|J_l(\hat{\Lambda}_d)J_{l+m}(\hat{\Lambda}_d)F_0(v_{\parallel,res})|$  from equation (\ref{eq:GAM_response_general}) with $v_{\parallel,res}=qR_0\omega_G/(l+m)$.   Deriving the GAM Landau damping rate  for short wavelength KGAM, which is preferentially excited via  DW interactions, then becomes challenging due to the non-trivial task of summing up all the relevant  transit harmonic resonances.
An alternative approach was developed in Ref. \citenum{FZoncaEPL2008}, which is equivalent to adding up all the transit harmonic resonances. Detailed derivations and interpretations were given in later publications \cite{ZQiuPPCF2009,LChenPoP2017}.  The anti-Hermitian part of the GAM dispersion function in the short wavelength limit ($k_r\rho_{ti}q^2\gg1$) is then given as
\begin{eqnarray}
D_I&=&\sqrt{2}\left[1-2\hat{b}+\frac{\omega^2_{dt}}{\omega^2}\left(1+\sqrt{2}\tau \left(\frac{7}{4}+\tau\right)\right)\right.\nonumber\\
&&\left.+\frac{\omega\omega^2_{tt}}{24}\left(-\frac{4}{\omega^3_{dt}}+\frac{\omega}{\omega^4_{dt}}\right)\right]\exp{\left(-\frac{\omega}{\omega_{dt}}\right)}\label{eq:GAM_kinetic_DR_imag}.
\end{eqnarray}
Note that, even though $D_I$ is proportional to $\exp(-\omega/\omega_{dt})$ and the leading order resonant particle response is $\delta H_{res}\propto 1/(\omega-\omega_d)$, the underlying resonant condition is not a 	``drift resonance". The wave-particle energy exchange  is due to  the summation of all the transit harmonic resonances, as clarified in great detail in Ref. \citenum{LChenPoP2017}.
The real frequency and collisionless damping rate of GAM, can then be derived from equations (\ref{eq:GAM_kinetic_DR_real}) and (\ref{eq:GAM_kinetic_DR_imag}), with FLR and FOW effects properly accounted for. The present  approach   to the wave-particle resonances in the short wavelength limit  has broad  applications in, e.g., EP anomalous transport by ITG DW \cite{WZhangPRL2008,ZFengPoP2013} and short wavelength EGAM excitation \cite{LChenPoP2017}.

The GAM dispersion relation can be modified by various effects,   such as  the connection length affected by   equilibrium magnetic  geometries  including aspect ratio \cite{ZGaoPoP2008,ZGaoPoP2010} and  elongation \cite{ZGaoPoP2008,ZGaoPoP2010,ABiancalaniPoP2017}, kinetic electrons \cite{LWangPPCF2011,HZhangPoP2010} and  $m=2$ electro-magnetic component due to finite $\beta\equiv 4\pi P_0/B^2_0$ \cite{LWangPoP2011,DZhouPoP2007,ASmolyakovPPCF2008,DZhouPoP2016,PAngelinoPoP2008}. The latter issue is connected with our analysis of the vorticity equation below, i.e., equation (\ref{vorticityequation}), where finite  electromagnetic component in the field line bending term (first term therein) comes from the curvature coupling term (third term therein), due to the  combined effect of geodesic curvature  and the up-down anti-symmetric density perturbation. Interested readers may refer to the original publications for  details.

\section{Energetic particle induced GAM: resonant excitation, global mode structure and nonlinear saturation}
\label{sec:EGAM}

Due to its finite real frequency, GAM can   resonate with EPs and be driven unstable by velocity space anisotropic  EPs if the EP resonant   drive is strong enough to overcome the   dissipations due to, e.g. thermal ion induced Landau damping and/or continuum damping. Since its observation in experiments  \cite{HBerkNF2006,RNazikianPRL2008,TIdoNF2015,LHorvathNF2016}, EGAM  has attracted attention due to its potential application as active control of DW turbulences \cite{DZarzosoPRL2013,RDumontPPCF2013,ZQiuEPS2014,DZarzosoNF2017}.  The theoretical interpretation was first given in Ref. \cite{GFuPRL2008}, taking an EP beam with slowing down distribution in energy and localized Gaussian in pitch angle.  The mechanism for EGAM drive, is  similar to the well-known beam-plasma instability (BPI) in a strongly magnetized plasma, where a positive energy plasma mode is coupled to a negative energy beam mode \cite{HBerkNF2010}. The local EGAM theory was  then generalized to different cases depending on  EP source drive \cite{DZarzosoPoP2012,HWangPRL2013,JBaptistePoP2014,DZarzosoNF2014,HWangPoP2015,MSasakiPoP2016,LChenPoP2017}. Worthwhile being mentioned   are    the sharp gradient in pitch angle induced by prompt loss  leading to fast EGAM  onset discussed in Ref. \cite{HBerkNF2010}, and a theory considering  not fully slowed down EP beam \cite{JCaoPoP2015} to explain the    EGAM experiments in Large Helical Device (LHD) with low collisionality \cite{TIdoNF2015}.

While the continuous spectrum is one of the key features of GAM   \cite{FZoncaEPL2008},  the  theories mentioned above on EGAM  ignored it by   focusing on deriving the local dispersion relation    \cite{HBerkNF2010,MSasakiPPCF2011,JCaoPoP2015}. Thus,   the associated radial structures, which were expected to play  important roles  in the EP linear and nonlinear dynamics \cite{ZQiuPST2011,FZoncaPPCF2015,FZoncaNJP2015}, were neglected.
The effect of GAM continuum on EGAM excitation was first pointed out in  \cite{FZoncaIAEA2008},
where, by matching across the singular resonant layer with the GAM continuous spectrum, a model dispersion relation of global EGAM was obtained, demonstrating the finite drive threshold  due to the GAM continuum damping and the similarity to energetic particle mode (EPM) \cite{LChenPoP1994}. The global properties of EGAM depend on the relative scale lengths of GAM continuum and EP density profile, and thus, on the coupling of EGAM to GAM continuum.
The excitation of EGAM by a radially localized EP beam was then investigated in \cite{ZQiuPPCF2010}. With the EP beam localized away from the position where the mode frequency matches that of the GAM continuum, the   continuum damping is minimized, and the obtained global EGAM radial mode structure shows that EGAM is self-trapped by the localized EP beam \cite{ZQiuPPCF2010,FZoncaPoP2000}, with an exponentially small tunneling coupling to propagating KGAM, resulting in an exponentially small EGAM excitation threshold.
The case of a radially broad EP beam     with a density profile scale length comparable with the characteristic scale length of GAM continuous spectrum was considered in Ref. \cite{ZQiuPoP2012}, which is more relevant to realistic tokamak conditions, and  the excited EGAM  is  shown to strongly couple to   GAM continuum, as expected.

In this Section, the major progresses in linear EGAM analytical theory are reviewed, with the local stability properties discussed in Sec. \ref{sec:EGAM_local}. The global EGAM theory \cite{ZQiuPPCF2010,ZQiuPoP2012} considering the EP profile and coupling to GAM continuum is presented in Sec. \ref{sec:EGAM_global}. Speculations on EGAM nonlinear saturation and EP transport are made in Sec. \ref{sec:EGAM_saturation}

\subsection{Local EGAM theory}\label{sec:EGAM_local}

In this section, the local EGAM theory will be discussed, with the case of the   slowing down distribution in energy and single pitch angle EP investigated in Sec. \ref{sec:EGAM_local_slowing_downing}, elucidating also  the similarity of EGAM to the well-known beam plasma instability.  In Sec. \ref{sec:EGAM_local_not_fully_slowing_down}, we will discuss the case with a not fully slowed down EP beam due to NBI in a plasma with low collisionality \cite{TIdoNF2015,JCaoPoP2015}; while the case with a sharp gradient in pitch angle due to prompt loss of injected neutral beam \cite{HBerkNF2010}  is discussed in Sec. \ref{sec:EGAM_local_prompt_loss}. These two cases   may relate  to the   fast onset of EGAMs in experiments \cite{JCaoPoP2015,HBerkNF2010}. In the analysis through out Sec. \ref{sec:EGAM},   small but finite $T_e/T_i$  is assumed,  such that  $\omega_{tr,e}\gg\omega_G$ and GAM/EGAM scalar potential  is dominated by $m=0$ component.  Note that,  despite the apparent  contradiction of this assumption with  LHD experimental observations \cite{TIdoNF2015} at high electron temperature, the theoretical analysis remains   qualitatively unchanged.

The EGAM equation is derived from the surface averaged quasi-neutrality condition
\begin{equation}
-\frac{e}{m_i}n_{0}k^2_r\frac{1}{\Omega^2_i}\left(1-\frac{\omega^2_G}{\omega^2}-\frac{G}{2} k^2_r\rho^2_{ti} \right)\overline{\delta\phi}_G+\overline{\delta
n_h}  = 0 \label{eq:QN_EGAM},
\end{equation}
with the thermal plasma response obtained in the previous sections,  $G$ is due to  thermal ion FLR/FOW effect, and its expression   is given in Ref. \cite{ZQiuPPCF2010} (equation (31) therein). The perturbed   EP density, $\overline{\delta n}_h$, is defined as
\begin{eqnarray}
\overline{\delta n_{h}}&=&2\pi
B_0\sum_{\sigma=\pm1}\int\frac{Ed\Lambda
dE}{|v_{\parallel}|}\overline{\left[\frac{e}{m}\frac{\partial F_{0,h}}{\partial E}\overline{\delta\phi}_G+J_G\delta H_h\right]},\nonumber\\
&&\label{eq:EP_density_perturbation}
\end{eqnarray}
with $\Lambda\equiv\mu/E$ denoting the particle pitch angle, and $\mu\simeq v^2_{\perp}/(2B_0)$ the magnetic moment.  The EP nonadiabatic response, $\delta H_{h}$, is given by the $m=0$ component of the general solution, equation (\ref{eq:GAM_response_general}), due to the  $T_e/T_i\ll1$ limit  assumed here
\begin{eqnarray}
 \delta H_{h}&=&-\frac{e}{m_i}\omega J_G\partial_EF_{0,h} \sum_p  \sum_l \nonumber\\
 &&\times\frac{i^{(p-l)} J_l(\hat{\Lambda}_{d,h})J_p(\hat{\Lambda}_{d,h})e^{i(l+p)\theta}\overline{\delta\phi}_{G}}{\omega-l\omega_{tr,h}}.\label{eq:EP_GAM_response_general} \end{eqnarray}

Note that, $k_r\equiv -i\partial_r$ is the   radial derivative operator, so equation (\ref{eq:QN_EGAM}) can be readily applied to study the global EGAM problem. 
In the local limit with  $\hat{\Lambda}_{d,h}\ll1$, i.e., the EP drift orbit width is much smaller than the characteristic wavelength of EGAM, the primary transit resonances $\omega=\pm\omega_{tr,h}$ dominate,  and thus, the optimal ordering for EGAM excitation is $T_h/T_i\sim q^2$. Keeping only $l=0,\pm1$ transit harmonics, and assuming well circulating EPs, one then has
\begin{eqnarray}
\overline{\delta n_{h}}
&=&A\int\frac{(2-\Lambda B_0)^2}{\sqrt{1-\Lambda
B_0}}\frac{B_0dEd\Lambda E^{5/2}\partial_EF_{0,h}}{2E(1-\Lambda
B_0)-\omega^2q^2R^2_0}.\nonumber\\
&& \label{eq:EGAM_deltan_h}
\end{eqnarray}
Here, $A=\sqrt{2}\pi ce^2k^2_r\overline{\delta\phi_G}/(B_0\Omega_i)$.

Equation (\ref{eq:QN_EGAM}), thus,  with perturbed EP density given by equation (\ref{eq:EGAM_deltan_h}) and thermal ion FLR effects neglected is  the general equation describing local EGAM excitation by well circulating EPs, with the specific cases characterized  by different equilibrium EP distribution function $F_{0,h}$.   EGAM excitation by  bounce resonance of deeply trapped EPs is investigated in Ref. \cite{IChavdarovski2017}, and will not be discussed here due to length constraints.

\subsubsection{Excitation by EP with slowing down distribution function}
\label{sec:EGAM_local_slowing_downing}

We start with the general case of EP distribution being slowing down in energy and localized in pitch angle \cite{GFuPRL2008,ZQiuPPCF2010}. This  reflects that EP collisions with thermal electrons (slowing down) are more efficient than that with ions (pitch angle scattering) at high EP velocity, and is consistent with the $\beta_h/\tau_{SD}\sim\beta_c/\tau_E$ ordering for plasma heated by EPs. Here, $\tau_{SD}$ is the typical slowing down time and $\tau_E$ is the energy confinement time.  This case was first investigated by Fu \cite{GFuPRL2008}, with the final eigenmode equation  (corresponding to equation (\ref{eq:QN_EGAM}) here) solved numerically to show that  the unstable branch is characterized by a frequency lower than  the local GAM frequency. Here,
in order to make further analytic progress, we take a single
pitch-angle slowing-down distribution for the EPs \cite{ZQiuPPCF2010}; i.e.,
$F_{0,h}=c_0(r)\delta(\Lambda-\Lambda_0)H_E$, where
  $c_0(r)=\sqrt{2(1-\Lambda_0B_0)}n_b(r)/(4\pi B_0\ln{(E_b/E_{c})})$, $n_b(r)$ is the density of the EPs beam,
  $E_b$ and $E_c$ are, respectively, the EP birth and critical energies \cite{TStixPoP1972}, $\delta(x)$ is the Dirac delta function,
   and $H_E=\Theta(1-E/E_b)/(E^{3/2}+E^{3/2}_{c})$, with $\Theta(1-E/E_b)$ being the Heaviside step function.
Noting that generally $E_b \gg E_c$, the local EGAM dispersion relation can be derived as:
\begin{eqnarray}
& & \mathscr{E}_{EGAM}=-1+ \omega^2_G/\omega^2+N_b [C  \ln(1-\omega^2_{tr,b}/\omega^2)  \nonumber \\
& &     + D (\omega_{tr,b}^2/\omega^2)/(1-\omega_{tr,b}^2/\omega^2) ] =
0;\label{eq:EGAM_local_SD}
\end{eqnarray}
where $\omega_{tr,b}=\sqrt{2E_b(1-\Lambda_0B_0)}/(qR_0)$ is the EP transit frequency at birth energy,
$N_b\equiv n_b\sqrt{1-\Lambda_0B_0}q^2/(4\ln{(E_b/E_{c})}n_c)\propto \beta_h$ (noting   $T_h/T_i\sim q^2$), $C\equiv (2-\Lambda_0B_0)(-2+5\Lambda_0B_0)/(2(1-\Lambda_0B_0)^{5/2})$ and $D=\Lambda_0B_0(2-\Lambda_0B_0)^2/(1-\Lambda_0B_0)^{5/2}$.

In equation (\ref{eq:EGAM_local_SD}), the first term in the EP response (i.e., the logarithmic term) corresponds to resonant EP drive and the second term contributes to frequency shift from local GAM continuum frequency. As a result, the EGAM instability requires $C>0$, i.e., 
\begin{eqnarray}
\Lambda_0B_0>2/5.
\end{eqnarray}

\begin{figure}
\includegraphics[width=7cm]{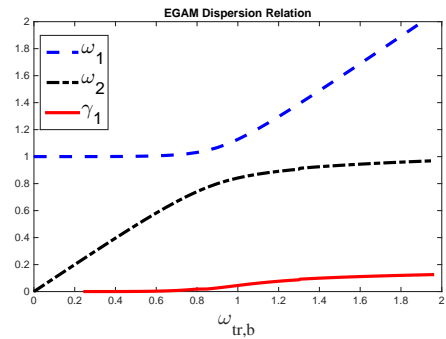}
\caption{EGAM dispersion relation}\label{fig:EGAM_DR}
\end{figure}

Equation (\ref{eq:EGAM_local_SD}) can be solved numerically, and the  numerical solution is shown in Fig. \ref{fig:EGAM_DR}. The real frequency and growth rate v.s. $\omega_{tr,b}$ are plotted in   units of $\omega_G$. It is shown that, when $\omega_{tr,b}$ is far away from $\omega_G$,  there are two branches with  frequency determined by GAM and $\omega_{tr,b}$, respectively. As $\omega_{tr,b}\simeq\omega_G$, these two branches are strongly coupled, and reconnect. The solid curve is the linear EGAM growth rate corresponding to the unstable branch $\omega_2$, the dot-dashed curve. The unstable mode frequency is always lower than the local GAM frequency \cite{GFuPRL2008}, consistent with experimental observations \cite{RNazikianPRL2008}.  The similarity of EGAM to the well-known BPI can be clearly seen from this figure. Note that the similarity of the EGAM in three dimensional tokamak to the BPI in a 1D strongly magnetized plasma is not coincidental.  The similarity lies in the fact that, due to the toroidally symmetry mode structure and low frequency, the toroidal angular momentum $P_{\phi}$ and mangetic moment $\mu$ are conserved, and EGAM is essentially quasi-1D with the dynamics only in $(J, \theta)$ space. Here, $J\equiv\int v_{\parallel} dl$ is the action conjugate to $\theta$ (second adiabatic invariant). This similarity  provides insights into,   not only the linear physics of EGAM, but also    EGAM nonlinear dynamics \cite{ZQiuPST2011,HWangPRL2013,ABiancalaniJPP2017} due to    wave-particle phase space nonlinear interactions.

\subsubsection{Excitation by not fully slowed down ion beam}
\label{sec:EGAM_local_not_fully_slowing_down}

The EGAM observed   in the Large Helical Device (LHD) \cite{TIdoNF2015}    during tangential neutral beam injection encountered some difficulties  in the comparison with  theoretical predictions \cite{GFuPRL2008,ZQiuPPCF2010}, because the EP birth energy ($\sim170KeV$) is much higher than that predicted for wave-particle resonance \cite{ZQiuPPCF2010}, and the observed EGAM frequency can be higher than local GAM frequency. The interpretation was given in Ref.  \cite{JCaoPoP2015}, noting that EGAM onset time is shorter than the slowing down time ($\tau_{SD}\sim 9 s$) of injected neutral beam due to the peculiar discharge condition with high temperature ($T_e\sim 7 KeV$), low plasma density ($n\sim0.1\times 10^{19}m^{-3}$) \cite{TIdoNF2015}. In that work \cite{JCaoPoP2015},
a local theory of EGAM excitation by a not fully slowed down EP beam is investigated. It is shown that   the instability drive comes from the positive velocity space gradient in the low-energy end of the EP distribution function \cite{JCaoPoP2015},    in addition to the velocity space anisotropy \cite{GFuPRL2008}.
For the sake of simplicity,   the helicity of the device is ignored and large aspect ratio is assumed,   consistent with the experimental observation in the center of the device using heavy ion beam probe \cite{TIdoNF2015}.
The EP distribution function is given as
\begin{eqnarray}
F_{0,h}=\frac{c_0H(E_b-E)H(E-E_L)}{E^{3/2}+E^{3/2}_{c}}\delta(\Lambda-\Lambda_0),\nonumber
\end{eqnarray}
which is  derived exactly from Fokker-Planck equation with only slowing down collisional operator, and an EP source term  with single pitch angle and birth energy. Here,  $c_0=\Gamma_b\tau_{SD}/(4\pi)$ with $\Gamma_b$ being the NBI particle flux,  $E_L\simeq E_b\exp(-2t/\tau_{SD})$ is the time dependent  lower energy end of the distribution function, and the modification of the EP distribution function due to interaction with EGAM is ignored.

The dispersion relation can then be derived as
\begin{eqnarray}
&&-1+ \omega^2_G/\omega^2+\pi B_0q^2 c_0/(\sqrt{2}n_0)\times\nonumber\\
&&\left[C\left(\ln(1-\omega^2_{tr,b}/\omega^2)-\ln(1-\omega^2_{tr,L}/\omega^2)\right)\right.\nonumber\\
&&+ \left.D\left(1/(1-\omega^2_{tr,b}/\omega^2)-1/(1-\omega^2_{tr,L}/\omega^2)\right)\right]=0.\nonumber\\
\end{eqnarray}
Here,   $\omega_{tr,b}$ and $\omega_{tr,L}$ are the transit frequencies defined at $E_b$ and $E_L$, respectively. Note that,  as discussed in the previous section  for the slowing down case,   the logarithmic singularity at $\omega_{tr,b}$ is destabilizing  given $\Lambda_0B_0>2/5$ and thus $C>0$, and  the simple pole at $\omega_{tr,b}$ will only contribute to modulate the EGAM frequency. However,  for the not fully slowed down distribution function,  considered here, the simple pole at $\omega_{tr,L}$ is also destabilizing and, thus,  there is no threshold in pitch angle.

The dispersion relation can be solved numerically as a function of $\tau=t/\tau_{SD}$, and yields the slow temporal evolution of the excited EGAM due to the slowing down of the EP beam.  $\omega_{tr,b}=3\omega_G$ is taken. There are three branches;  a GAM branch with $\omega_r\simeq \omega_G$, a lower beam branch (LBB) with $\omega_r\simeq \omega_{tr,L}(t)$; and an upper beam branch(UBB), with $\omega\simeq \omega_{tr,b}$.

\begin{figure}
\includegraphics[width=7cm]{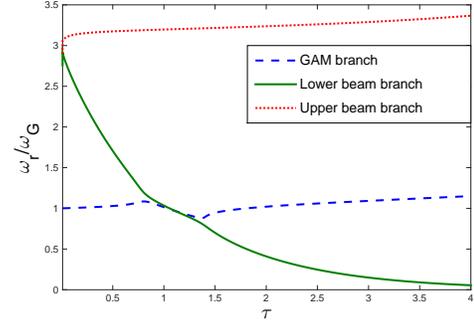}
\caption{(Reproduced from Fig. 3 of Ref. \citenum{JCaoPoP2015}.) Real frequency for $\Lambda_0B_0<2/5$}\label{fig:RF_SPA}
\end{figure}

\begin{figure}
\includegraphics[width=7cm]{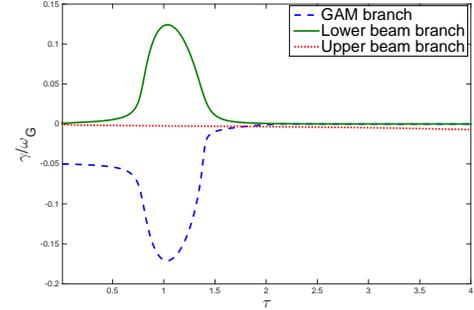}
\caption{(Reproduced from Fig. 4 of Ref. \citenum{JCaoPoP2015}.)   Growth rate for $\Lambda_0B_0<2/5$}\label{fig:GR_SPA}
\end{figure}

The real frequency and growth rate for $\Lambda_0B_0<2/5$  are shown in Figs. \ref{fig:RF_SPA} and \ref{fig:GR_SPA}, respectively.  We can see that,  only the LBB is unstable.  In this case, the logarithmic term is stabilizing \cite{ZQiuPPCF2010}; thus,  the EGAM discussed here is similar to BPI, which, however, has a double pole instead of the simple pole as in the present case.   However, when $\omega_{tr,L}$ becomes smaller than $\omega_G$ by a finite amount, the growth rate of LBB decreases to zero as the contribution of the simple pole becomes vanishingly small, similar to that of BPI. The strong instability at $\omega_L(t)\simeq\omega_G$ may   provide an explanation for the fast growth of  EGAM observed experimentally.
We also note that  the frequency of the unstable LBB  can be significantly larger than $\omega_G$, as is shown in Fig. \ref{fig:RF_SPA}.  This may explain the higher-frequency branch of EGAM observed in LHD \cite{TIdoNF2015}.

\begin{figure}
\includegraphics[width=7cm]{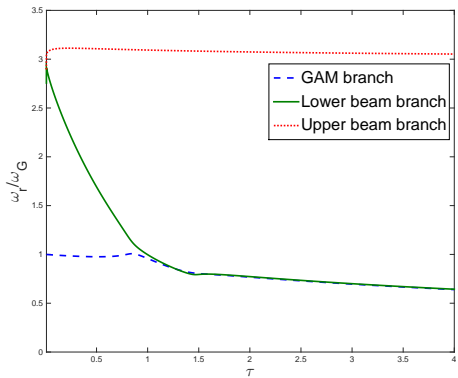}
\caption{(Reproduced from Fig. 1 of Ref. \citenum{JCaoPoP2015}.) Real frequency for $\Lambda_0B_0>2/5$}\label{fig:RF_LPA}
\end{figure}

\begin{figure}
\includegraphics[width=7cm]{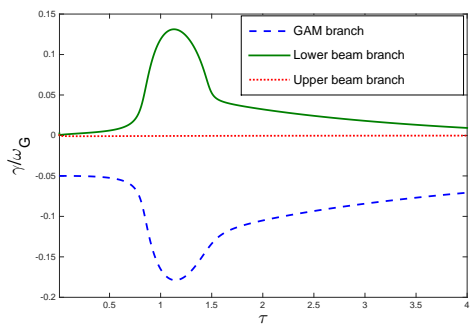}
\caption{(Reproduced from Fig. 2 of Ref. \citenum{JCaoPoP2015}.) Growth rate for $\Lambda_0B_0>2/5$}\label{fig:GR_LPA}
\end{figure}

On the other hand, for  $\Lambda_0B_0>2/5$, i.e., $C>0$, the real frequencies and growth rates are shown, respectively, in Figs. \ref{fig:RF_LPA} and \ref{fig:GR_LPA}. The EGAM problem can be understood as a double-beam plasma instability, with the two singularities (logarithmic singularity at $\omega_{tr,b}$ and simple pole at $\omega_{tr,L}$) contributing at different values of $\omega_{tr,L}/\omega_G$. 
The major difference with the previous case with $\Lambda_0B_0<2/5$ is that, as $\omega_{tr,L}$ further decreases ($\tau>1.5$), the growth rate decays very slowly, due to the contribution of the destabilizing logarithmic term.

Note that  a similar explanation was given in Ref. \cite{HWangPoP2015}, assuming a positive slope in the distribution function due to finite charge exchange time.   The interpretation given here, with slight modification to the one discussed in Sec. \ref{sec:EGAM_local_slowing_downing},  can recover all the peculiar features of the LHD EGAM experiment, and the theory can be applicable to potential experimental results obtained  from devices with similar features, for example  EAST \cite{JLiNP2013}.

\subsubsection{Fast EGAM onset due to  sharp gradient in pitch angle induced by prompt loss} 
\label{sec:EGAM_local_prompt_loss}

In DIII-D experiments,  EGAM was excited by tangential NBI with relatively large pitch angle \cite{RNazikianPRL2008}, and EGAM was observed in less than a millisecond after the turn-on of NBI \cite{HBerkNF2010}. A new mechanism was then proposed to explain the fast excitation based on the prompt loss induced sharp gradient in pitch angle, which can explain why modes were often observed during  counter-injection \cite{HBerkNF2010}.  Here,  the main steps of the theory will be briefly summarized, while  interested readers may refer to the original paper \cite{HBerkNF2010} for detailed derivations.

In Ref. \cite{HBerkNF2010}, the EPs were generated by  NBI with a single injection velocity $u_0$ and    pitch angle $\Lambda$ narrowly localized around $\Lambda_0$. After one transit/bounce time, the unconfined barely trapped particles with $\Lambda B_0\sim 1$ were lost, leaving a sharp gradient in the pitch angle, and the effective EP distribution could be modelled as
 \begin{eqnarray}
&&F_{0,h}=St \frac{\delta (u-u_0)}{u^2_0}\frac{3}{8\pi\Delta\Lambda}\left(1-\left(\frac{\Lambda-\Lambda_0 }{\Delta\Lambda}\right)^2\right)\nonumber\\
&\times&\Theta(\Lambda-\Lambda_0+\Delta\Lambda)\Theta(\Lambda_0-\Lambda+\Delta\Lambda)\Theta(\Lambda-\Lambda_c),
\end{eqnarray}
with the pitch angle  $\Lambda\equiv u_{\parallel}/u$,  used only in this subsection  to be consistent with the original paper,  $\Delta\Lambda$ denoting  the spreading of pitch angle,  $\Lambda_c$ is the loss boundary,  and $\Lambda_0-\Delta\Lambda<\Lambda_c<\Lambda_0+\Delta\Lambda$. Furthermore, $S$ is the NBI particle flux   and $St$ is the EP density.

Noting that
$\partial_E|_{\mu}  F_{0,h}=\left(\partial_E|_{\Lambda} +(\partial_{E}\Lambda)\partial_{\Lambda}|_E \right)F_{0,h}$, the sharp gradient at $\Lambda_c$ may  induce strong EGAM drive, and the time needed for the  building up of the sharp gradient is of order $\sim \omega^{-1}_{tr,h}$, i.e., one transit/bounce period of the barely trapped EPs.  In Ref. \cite{HBerkNF2010}, the EGAM dispersion relation was derived, and solved perturbatively for the beam mode \cite{HBerkNF2010,ZQiuPPCF2010}. It was found that, with the existence of sharp gradient, the EGAM onset time was very short, and can be applied to interpret the DIII-D results \cite{RNazikianPRL2008}. The drive was strongest as $\Lambda_0=\Lambda_c$, i.e., the  NBI    was maximized at the loss boundary, producing an EP density maximized at the discontinuity of the distribution function.

In the treatment of Ref. \cite{HBerkNF2010}, however, the GAM Landau damping or other possible dissipation channels are missing,  which is usually not important for EGAM local instabilities with a given EP density, since GAM Landau damping can be weak compared  to the EP resonant drive. However, in the case considered here for the ``fast onset" of EGAMs with EPs density accumulating with time, a finite dissipation may induce a finite threshold on EP density ($n_{cr}$),  and it takes $\tau_c\sim n_{cr}/S$ for the EP density to accumulate.  For EGAM with a finite linear growth rate as EP density is above the threshold of marginal instability, the onset time will be $\max(\tau_c, 1/\gamma_L)$  with $\gamma_L$ being the obtained EGAM linear growth rate.

Besides the cases reviewed above \cite{GFuPRL2008,ZQiuPPCF2010,JCaoPoP2015,HWangPoP2015,HBerkNF2010}, other EP equilibrium distributions were considered, including bump-on-tail \cite{JBaptistePoP2014,ABiancalaniNF2014,DZarzosoPoP2012,DZarzosoNF2014}, and a careful examination of beam v.s. GAM branch was carried out  \cite{DZarzosoPoP2012,JBaptistePoP2014}. Corrections to EGAM local dispersion relation due to electromagnetic effects \cite{LWangPoP2014},  kinetic electrons \cite{ABiancalaniPoP2017,ABiancalaniEFTC2017} and toroidal rotation \cite{HRenNF2017}, were also investigated.    Readers interested in these works may refer to the original papers for more details.

\subsection{Global theory}
\label{sec:EGAM_global}

EGAM may have a global mode structure due to the coupling to GAM continuum, and the nonlocal properties of EGAM are determined by  the relative orderings of two scale lengths, i.e., the characteristic scale length of GAM continuous spectrum $L_G\equiv |\omega^2_G(r)/(\partial\omega^2_G(r)/\partial r)|$ and the scale length of EP density profile $L_E\equiv |n_{0,h}(r)/(\partial n_{0,h}/\partial r)|$.

In the absence of GAM continuum, EGAM can be self-trapped by EP density profile, and form a radial EGAM eigenstate \cite{ZQiuPPCF2010,FZoncaPoP2000} with a radial scale length of $\sqrt{\rho_{d,h}L_E}$, as we will discuss in Sec. \ref{sec:EGAM_trapping}.
Noting that  the EGAM frequency can be significantly  lower than local GAM frequency due to non-resonant EP effects \cite{GFuPRL2008,ZQiuPPCF2010}, the EGAM coupling to GAM continuum can be minimized by localizing the driving EP beam away from   where EGAM frequency matches the local GAM frequency, given $L_E\ll L_G$ \cite{ZQiuPPCF2010}.   In this limit, the exponentially small  tunnelling coupling to KGAM will lead to a threshold condition on EGAM excitation \cite{ZQiuPPCF2010}. However, for more realistic cases with $L_E$ comparable to $L_G$, the EGAM will strongly  couple to GAM continuum \cite{FZoncaIAEA2008}, leading to a higher threshold on EGAM excitation \cite{ZQiuPoP2012}.  In this subsection, the global feature of EGAM will be discussed, for different  $L_E/L_G$ such that EGAM coupling to GAM continuum are, respectively, vanishing  ($L_E/L_G=0$), weak ($L_E/L_G\ll1$) and strong ($L_E/L_G\sim O(1)$).

\subsubsection{Radially localized EP drive: EGAM radial eigenstate}\label{sec:EGAM_trapping}

We start with EGAM excitation by a radially  localized EP beam in uniform  thermal plasmas.  To account for the global features,  kinetic effects should be included to obtain the global mode equation, and the EP FOW effects dominate.  Noting that $k_r=-i\partial_r$,  the EGAM mode equation can be written as
\begin{eqnarray}
\left[\frac{\partial}{\partial r}\left(-\frac{1}{2}\rho^2_{d,b}N_b(r)H\right)\frac{\partial}{\partial r}+\mathscr{E}_{EGAM}(r) \right]\delta E=0,\label{eq:EGAM_local_trapping}
\end{eqnarray}
with $H\sim O(1)$ due to  EP FOW effects, and its expression  given in equation (21) of Ref. \cite{ZQiuPPCF2010}, and $\rho_{d,b}=\rho_{d,h}(E_b,\Lambda_0)$.  The characteristic scale length of the mode is $\Delta\simeq \sqrt{\rho_{d,b}L_E} \ll L_E$ to be shown  a posteriori. Expanding $N_b(r)\simeq N_b(r_0)(1-(r-r_b)^2/L^2_E)$ and introducing $r-r_b=\xi z$,  the mode equation  becomes
\begin{equation}
\left[\frac{\partial^2}{\partial
z^2}+2\xi^2\left[-\frac{\mathscr{E}_{EGAM}(r_b)}{\rho^2_{d,b}N_b(r_b)H}\right]-z^2\right]\delta
E=0, \label{eq:EGAMloc}
\end{equation}
where $\xi^4=\rho^2_{d,b}N_b(r_b)HL^2_E/(2(-1+\omega^2_G(r_b)/\omega^2))$
and causality constraint must be applied in determining $\xi^2$.
Equation (\ref{eq:EGAMloc}) is the typical  Weber equation and its eigenvalues
satisfy the following ``localized" EGAM dispersion relation
(i.e., neglecting the coupling to the GAM continuum)
\begin{equation}
2\xi^2\left[-\frac{\mathscr{E}_{EGAM}(r_b)}{\rho^2_{d,b}N_b(r_b)H}\right]=2l+1,
\ \ l=0,1,2,\cdots. \label{locdisp}
\end{equation}
Here, $l$ is the radial eigenmode number. Meanwhile, the radial
electric field is
\begin{equation} \delta E\propto
H_l((r-r_b)/\xi)\exp(-(r-r_b)^2/(2\xi^2)),\label{innermodestructure}
\end{equation}
with $H_l$ being the $l-th$ Hermite polynomial. The ground state with $l=0$ is the most unstable mode with the straightforward interpretation as the mode structure localized at strongest EP drive.

\subsubsection{Radially localized EP drive: Weak  tunneling coupling to GAM continuum}
\label{sec:EGAM_global_weak}

\begin{figure}
 \includegraphics[width=3.0in]{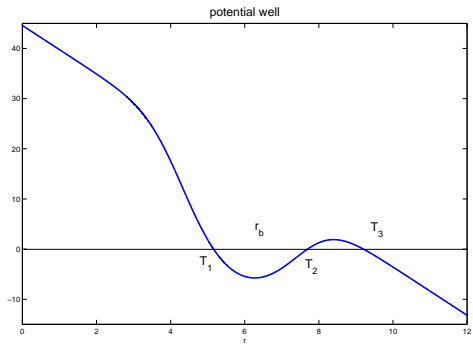}
\caption{(Reproduced from Fig. 2 of Ref. \citenum{ZQiuPPCF2010}.) Potential well: $-Q$ vs $r/L_b$.} \label{fig:potentialwell}
\end{figure}

Considering weak but finite thermal temperature gradient  with $L_G\gg L_E$, the EGAM can be  coupled to GAM continuum at the point the EGAM frequency matches the local GAM frequency,  and the coupling is weak since EGAM mode amplitude is exponentially small at the resonance point. Note that, although thermal ion FLR/FOW is formally  much smaller than
 EP FOW, kinetic effect is dominated by thermal ion FLR as EP density diminishes.  Noting that the typical scale length of EGAM is $L\simeq \sqrt{\rho_{d,b}L_E}\ll L_E, L_G$, the mode equation can be written as
\begin{equation}
\left[\partial^2_r+Q(r)\right]\delta
E=0\label{eq:EGAMsch},
\end{equation}
with $-Q(r)=2\mathscr{E}_{EGAM}(r)/(\rho^2_{d,b}N_b(r)H+\rho^2_{ti}G)$ being the  potential well.  The kinetic dispersiveness amplitude is given by $\rho^2_{d,b}N_b(r)H + \rho^2_{ti}G$, with the first term due to  EP FOW while the second term accounts for thermal ion FLR/FOW, and the expression of $G$ was given in Ref. \citenum{ZQiuPPCF2010} (equation (31) therein). In the EP localization region, kinetic dispersiveness is dominated by EP FOW, and we recover equation (\ref{eq:EGAM_local_trapping}); while, as EP fade  away, equation (\ref{eq:EGAMsch}) reduces to that describing KGAM propagation:
\begin{equation}
[\partial^2_r+2\left(1-\omega^2_G(r)/\omega^2\right)/(\rho^2_{ti}G)]\delta
E=0, \label{eq:EGAM_mc_kgam}
\end{equation}
and the KGAM radial electric field exhibits the  characteristic   Airy scale $k_r\sim O(\rho^{2/3}_{ti}L^{1/3}_G)$.  The potential well, $-Q(r)$, is given by Fig. \ref{fig:potentialwell}, with three regular turning points (zeros), $T_1$,
$T_2$ and $T_3$.   $T_1$ and $T_2$ are
the turning points pair due to the localization effect of EPs,  and
form a bound state as we have discussed for
equation~(\ref{eq:EGAM_local_trapping}). $T_3$ is the turning point accounting for mode
conversion to KGAM, beyond which the mode propagates
outward, as noted in the discussion following equation~(\ref{eq:EGAM_mc_kgam}).

Away from the turning points, $Q(r)$ is slowly varying and equation (\ref{eq:EGAMsch}) can  be solved using WKB approach. In particular, we obtain
\begin{eqnarray}
\delta
E&=&\frac{1}{Q^{1/4}(r)}\left[A_1\exp(i\int\sqrt{Q(r)}dr)\right.\nonumber\\
&&\hspace*{3.5em}\left.+B_1\exp(-i\int\sqrt{Q(r)}dr)\right]\label{eq:WKBsolution}.
\end{eqnarray}

\begin{figure}
\includegraphics[width=3.0in]{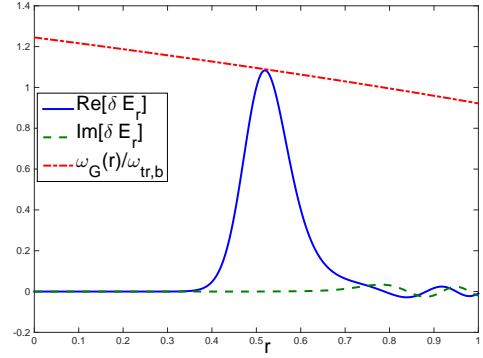}
\caption{(Reproduced from Fig. 4 of Ref. \citenum{ZQiuPPCF2010}. ) Sharply distributed EP: global mode structure.} \label{Fig:EGAM_global_tunnel}
\end{figure}

\begin{figure}
\includegraphics[width=3.0in]{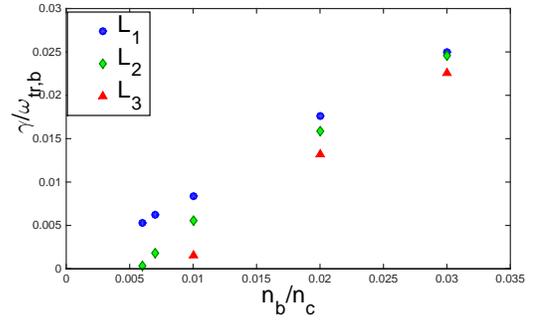}
\caption{(Reproduced from Fig. 3 of Ref. \citenum{ZQiuPPCF2010}. ) Sharply distributed EP: EGAM excitation threshold increases with decreasing $L_G$.} \label{Fig:EGAM_weak_threshold}
 \end{figure}

The corresponding WKB dispersion relation of the eigenmode described
by equation (\ref{eq:EGAMsch}) can then be straightforwardly derived
via asymptotic  matching of the WKB solutions, equation
(\ref{eq:WKBsolution}), across the turning points and is given by
\begin{equation}
e^{2 i W_1}=(e^{2 i W_2}+1)/(e^{2 i W_2}-1); \label{globdisp}
\end{equation}
where $W_1=\int^{T_2}_{T_1}\sqrt{Q(r)}dr$ and
$W_2=\int^{T_3}_{T_2}\sqrt{Q(r)}dr$. The tunneling coefficient
$e^{2iW_2}$ is formally exponentially small, and the WKB eigenmode
dispersion relation of EGAM becomes approximately
\begin{equation}
W_1=(l+1/2)\pi-ie^{2iW_2},\ \ \ \ \ \
\mbox{l=0,1,2,$\cdots$}.\label{eq:simglobdisp}
\end{equation}
Equation ~(\ref{eq:simglobdisp}) is the well-known
Bohr-Sommerfeld quantization condition including the tunneling
coupling to outgoing KGAM.  Neglecting the tunneling coupling in the $L_E/L_G\rightarrow\infty$ limit,  equation (\ref{eq:simglobdisp}) is equivalent to  equation (\ref{eq:EGAMloc}).  Near marginal stability,   the global EGAM growth rate can be obtained from the imaginary part of equation (\ref{eq:simglobdisp})
\begin{equation}
\gamma=-W_{1i}/(\partial W_{1r}/\partial\omega_r)-e^{2 i
W_2}/(\partial W_{1r}/\partial\omega_r);\label{gamma}
\end{equation}
expressing the mode excitation when the EP resonant drive exceeds the tunneling-convective damping, and $\omega_r$ is solved from $W_{1r}(\omega_r)=0$, where $W_{1r}$ and $W_{1i}$ are, respectively, the real and
imaginary parts of $W_1$ \cite{ZQiuPPCF2010}. The mode structure of EGAM from numerical solution of equation~(\ref{eq:EGAMsch}) (cf. Fig. \ref{Fig:EGAM_global_tunnel}) shows mode trapping by localized EP drive   with an exponentially small tunneling of the electric field to an outward propagating KGAM due to coupling to  GAM continuous spectrum,    and it is very similar to the DIII-D observations by Nazikian et al \cite{RNazikianPrivate2009}. Meanwhile, the EGAM threshold condition, due
to non-local coupling to  KGAM, is expected to
increase for decreasing $L_G$, and is shown numerically in Fig. \ref{Fig:EGAM_weak_threshold} for $L_G=L_1,L_2,L_3$ with $L_3<L_2<L_1=\infty$.

\subsubsection{Radially broad EP drive: Strongly coupling to GAM continuum}
\label{sec:EGAM_global_strong}

It is shown in Figs. \ref{Fig:EGAM_global_tunnel} and \ref{Fig:EGAM_weak_threshold} that, the EGAM coupling to GAM continuum increases as  its mode radial width $\propto\sqrt{\rho_{d,h} L_E}$ increases with respect to $L_G$.
In realistic tokamak plasmas,  it is expected that the EP  density profile  scale length $L_E$ is comparable to $L_G$  and, thus, the excited EGAM is expected to be strongly dependent on the radial mode structure determined by both radial profiles of EP drive and GAM continuum. As a result, the
 normalized  EP drift orbit, $k_r\rho_{d,h}$, changes continuously due to the change of $k_r$. Away from the singular point,   EGAM   is characterized by regular radial structure, with $k_r\rho_{d,h}\simeq\sqrt{\rho_{d,h}/L_E}\ll1$, as we discussed in Sec. \ref{sec:EGAM_trapping} and \ref{sec:EGAM_global_weak}. At the resonant  coupling position to GAM continuum, however, the mode structure is characterized by $k_r\sim   \rho^{-2/3}_{ti}L^{-1/3}_G$, considering the singularity is removed by thermal ion FLR effects, and EPs respond adiabatically to the mode  ($|k_r\rho_{d,h}|\gg1$). In between the regular region and singular layer, the EGAM wavelength varies continuously, and the EGAM eigenmode equation is an integral-differential equation, which generally requires numerical solution.

\begin{figure}
 \includegraphics[width=3.5in]{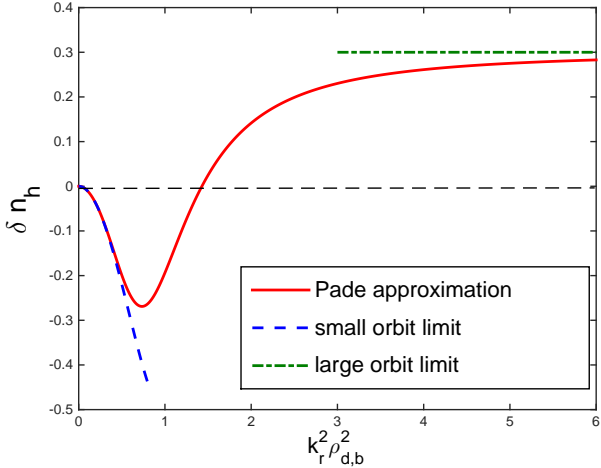}
\caption{(Reproduced from Fig. 1 of Ref. \citenum{ZQiuPoP2012}.) Pad\'e approximation surface averaged perturbed EP density. In which,  the solid curve is the Pad\'e approximation of the EP response,  while the dashed   and the dotted-dashed curves are  the small and large drift orbit widths responses, respectively.} \label{fig:padenh}
 \end{figure}

In Ref. \citenum{ZQiuPoP2012}, the EP response is modelled by Pad\'e approximation, which recovers  the EP response at $k_r\rho_{d,h}\ll1$ and $k_r\rho_{d,h}\gg1$ limit, and varies continuously with $k_r\rho_{d,h}$:
\begin{eqnarray}
\overline{\delta n_h}&=&\frac{e}{m}\frac{n_cN_b\overline{\delta\phi}}{\Omega^2_i}k^2_r\frac{\mathscr{E}_{h0}+Hk^2_r\rho^2_{d,b}/2}{1+k^4_r\rho^2_{d,b}n_cN_bH/(2n_E\delta\mathscr{E}_h)}\nonumber\\&\equiv&\frac{e}{m}\frac{n_cN_b\overline{\delta\phi}}{\Omega^2_i}k^2_r\mathscr{V}(k_r).\label{eq:EGAM_EP_Pade}
\end{eqnarray}
Therefore, this Pad\'e approximation EP response,  as shown in Fig. \ref{fig:padenh},   asymptotically captures  the EP response as $k_r\rho_{d,h}$ varies. We note, here, that   the equivalent potential function $\mathscr{V}(k_r)$ is independent of $r$.

Taking a linear radial dependence of the   GAM dielectric function, $\mathscr{E}_c\simeq\mathscr{E}_{c0}(1-(r-r_0)/L_G)$,  and assuming a Lorentian distribution for the EP radial density profile, $n_E(r)=n_E(r_0)/(1+(r-r_0)^2/L^2_E)$, the EGAM eigenmode equation is reduced to a third order differential equation in the Fourier space, i.e., 
\begin{eqnarray}
&&\left[\left(\frac{i}{L_G}\frac{\partial}{\partial k_r}-1\right)\left(\frac{\partial^2}{L^2_E\partial k^2_r}-1\right)\right.\nonumber\\
&&\hspace*{6em}\left.+\frac{N_b(r_0)}{\mathscr{E}_{c0}}\mathscr{V}(k_r)\right]\delta E_r=0. 
\label{eq:EGAM_Pade_reduced}
\end{eqnarray}\normalsize
Note that, in equation (\ref{eq:EGAM_Pade_reduced}), kinetic effects associated with thermal ion FLR  are neglected by taking $G=0$ since the mode equation in Fourier-$k_r$ space is regular; consequently, the contribution of GAM continuum in the reduced equation on EGAM excitation is   continuum  damping instead of mode conversion \cite{GVladRNC1999}.

As $|k_r\rho_{d, h}|\rightarrow\infty$,  $\mathscr{V}(k_r)$ vanishes as $O(1/k^2_r)$, and      equation (\ref{eq:EGAM_Pade_reduced})   has the following (out-going wave) boundary condition:
\begin{eqnarray}
\delta E_r(k_r\rightarrow+\infty)&=&\hat{A}\exp(-L_Ek_r)+\hat{B}\exp(-iL_Gk_r), \nonumber\\
\delta E_r(k_r\rightarrow-\infty)&=&\hat{C}\exp(L_Ek_r),\nonumber
\end{eqnarray}
with the two exponentially decay terms   reflecting the fact that   EGAM cannot be effectively driven at small radial scales with $k_r\rho_{d,h}\gg1$; while the $\exp(-iL_Gk_r)$ term,   with a positive (outward) ``group velocity'' in Fourier space,  corresponds to generation of singular radial mode structures at the resonant point with GAM continuum and resulting into finite continuum damping. If the thermal ion FLR/FOW effect is properly taken into account,  it creates an additional potential well \cite{GVladRNC1999}   and prevents the mode structure in Fourier space to propagate into regions with $|k_r|\gg\rho^{-2/3}_{ti}L^{-1/3}_G$. This effect,  of course,  corresponds to resolving the singularity in real space   and describes thus,  mode conversion to kinetic GAM \cite{AHasegawaPoF1976,FZoncaEPL2008} due to thermal ion FLR effects.

The analytic  dispersion relation of the reduced Pad\'e EGAM eigenmode equation, equation (\ref{eq:EGAM_Pade_reduced}), can be formally derived via a variational principle.  Multiplying $\delta E^*_r$ to equation (\ref{eq:EGAM_Pade_reduced}), subtracting its complex conjugate,   and  integrating over the Fourier space,
we then get the formal dispersion relation of the global EGAM:
\begin{eqnarray}
&&\gamma\int^{\infty}_{-\infty}N_b(r_0)\frac{\partial Re(\mathscr{V}(k_r)/\mathscr{E}_{c0})}{\partial \omega_r}|\delta E_r|^2dk_r\nonumber\\
&=&-\int^{\infty}_{-\infty}N_b(r_0)Im\left(\frac{\mathscr{V}(k_r)}{\mathscr{E}_{c0}}\right)|\delta E_r|^2dk_r\nonumber\\
&&+\frac{(L^2_G+L^2_E)\hat{B}^2}{2L_GL^2_E}. \label{eq:EGAM_global_DRformal}
\end{eqnarray}

In equation (\ref{eq:EGAM_global_DRformal}),the left hand side represents the   rate of change of the total energy  and $\gamma$ is the imaginary part of eigenmode frequency $\omega$. On the right hand side, the first term represents the EP resonant drive,  while the second term represents   dissipation due to generation of short wavelength structures, i.e., continuum damping; and ``$\hat{B}$'' corresponds to the ratio of the mode amplitude at the resonant point compared  to that at the center of EP localization region,  and is to be determined from numerical solution of the reduced EGAM eigenmode equation.    Thus, equation (\ref{eq:EGAM_global_DRformal})  is exactly the Fourier space counterpart of equation (11) of Ref. \citenum{FZoncaIAEA2008}, describing  the EGAM excitation as  EP drive  in the ideal region exceeds the threshold  due to continuum damping  in the inertial layer, analogous to the well studied EPM problem,  including fishbone \cite{LChenPoP1994,LChenPRL1984,LChenRMP2016,FZoncaPoP2014b}.

 \begin{figure}
 \includegraphics[width=3.1in]{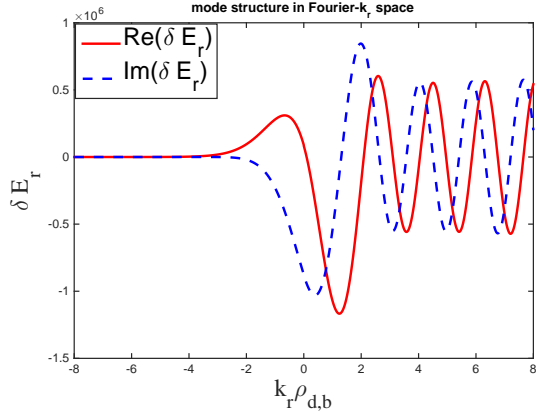}
\caption{(Reproduced from Fig. 2 (left panel) of Ref. \citenum{ZQiuPoP2012}.)  EGAM eigenmode structure in Fourier-$k_r$ space with $L_G/L_E=3.5$. The solid and dashed curves are respectively the real and imaginary part of the perturbed radial electric field.  } \label{fig:EGAM_mode_fourier}
 \end{figure}

  \begin{figure}
 \includegraphics[width=3.1in]{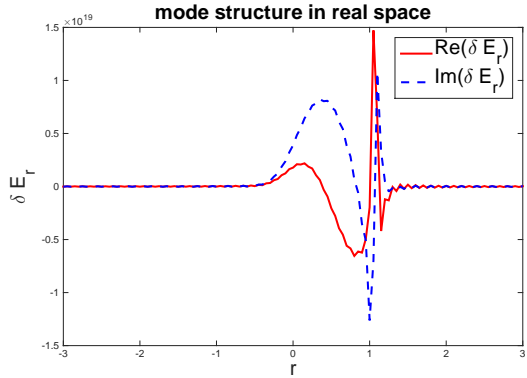}
\caption{(Reproduced from Fig. 2 (right panel) of Ref. \citenum{ZQiuPoP2012}.)  EGAM eigenmode structure in real space of the same case as Fig. \ref{fig:EGAM_mode_fourier}.} \label{fig:EGAM_mode_real}
 \end{figure}

  \begin{figure}
 \includegraphics[width=3.5in]{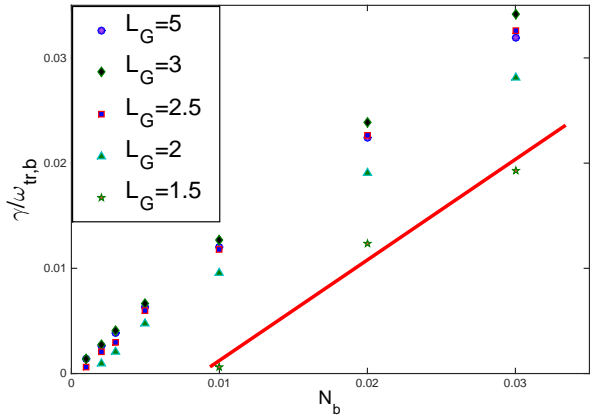}
\caption{(Reproduced from Fig. 3 of Ref. \citenum{ZQiuPoP2012}.)  Dependence of EGAM growth rate on normalized EP density for different $L_G/L_E$. The solid line is a linear fitting of the stars corresponding to $L_G/L_E=1.5$. } \label{fig:EGAM_global_threshold}
 \end{figure}

Equation (\ref{eq:EGAM_Pade_reduced}) is solved using a shooting code, and the obtained  structure of the most unstable mode is given in Fig. \ref{fig:EGAM_mode_fourier}, which is localized at small $k_r$ with a long tail to large $k_r$.    Note that,  although  Fig. \ref{fig:EGAM_mode_fourier} looks very similar to Fig. \ref{Fig:EGAM_global_tunnel} for localized EP drive,
  physics picture is in fact very different. Figure \ref{Fig:EGAM_global_tunnel} is the real space mode structure in the $L_E/L_G\ll1$ limit,  and the long tail corresponds to mode conversion to KGAM due to thermal ion FLR effects at the point EP density becomes vanishingly small. Figure \ref{fig:EGAM_mode_fourier}, meanwhile, shows  the Fourier space mode structure in the $L_E/L_G\sim O(1)$ limit, and the long tail corresponds to singular mode structure due to coupling to GAM continuum where EP density is finite, and thermal ion FLR effect is neglected.
By  Fourier transformation of fluctuation structures in Fig. \ref{fig:EGAM_mode_fourier}, the corresponding mode structure in real space is  given in Fig. \ref{fig:EGAM_mode_real}, and the significant difference with respect to Fig. \ref{Fig:EGAM_global_tunnel} becomes obvious.  
The increasing threshold on EP density  due to strong coupling to GAM continuum is shown in Fig. \ref{fig:EGAM_global_threshold}, as we have anticipated.

Note that, in  Ref. \citenum{ZQiuPoP2012}, the Pad\'e approximation of EP response  captures the feature of EP drive dependence on EGAM radial wavenumber,  while   the resonant drive is from the primary transit resonance $\omega=\pm\omega_{tr,h}$. This is qualitatively valid since the global mode structure is emphasized here. As we discussed in Sec. \ref{sec:GAM_kinetic_theory},  in the short wavelength limit with $k_r\rho_d\sim O(1)$, higher order transit harmonic resonances will also contribute and significantly increase wave-particle energy exchanges \cite{HSugamaJPP2006,XXuPRL2008,ZQiuPPCF2009,ABiancalaniPoP2017,LChenPoP2017}. It would be interesting to have the general integral-differential equation  with realistic EP response given as equation (\ref{eq:EP_GAM_response_general})  solved numerically, and compared to the  results based on the Pad\'e approximation of EP response discussed here.

 \subsection{Nonlinear EGAM saturation and EP transport}
 \label{sec:EGAM_saturation}

The nonlinear dynamics of EGAM can be understood using the analogy of EGAM with the one dimensional BPI, as we discussed in Sec. \ref{sec:EGAM_local_slowing_downing}.  The nonlinear evolution of EGAM, due to the nonlinear interactions with EPs,  can be obtained from equation (\ref{eq:QN_EGAM}), with the perturbed EP response derived from equation (\ref{eq:EGAM_deltan_h})   and  the evolution of the  ``equilibrium" EP distribution function, $F_{0,h}$, due to nonlinear interactions with EGAM properly taken into account. The $F_{0,h}$  evolution   due to nonlinear interaction with EGAM,  can be shown to obey   the following Dyson equation \cite{FZoncaNJP2015,MKakubook1993,FZoncaPPCF2015} 
\begin{eqnarray}
\bar{\omega}\hat{F}_{0,h}(\bar{\omega})&=&-\frac{e^2\hat{\omega}_{dr}}{16}|\delta\phi_G|^2\frac{\partial }{\partial E}\left[\frac{\hat{\omega}_{dr}(\bar{\omega}-i\gamma)}{(\bar{\omega}-i\gamma)^2-(\omega^2_{0r}-\omega_{tr})^2}\right]\nonumber\\
&\times&\frac{\partial}{\partial E}\hat{F}_{0,h}(\bar{\omega}-2i\gamma) +i F_{0,h}(0)\label{eq:EP_dyson}.
\end{eqnarray}
Here,  $\hat{F}_{0,h}$ is the Laplace transform of $F_{0,h}$, $\bar{\omega}$ denotes the slow  nonlinear time scale for $F_{0,h}$ evolution from its initial value $F_{0,h}(0)$,  and $|\gamma|\ll\omega_{0r}$ is the  growth rate of $\delta\phi_G$. Equation (\ref{eq:EP_dyson}) is of the form   of a Dyson equation, and describes the evolution of  $F_{0,h}$, due to emission and reabsorption of a single coherent EGAM. Note that, in deriving equation (\ref{eq:EP_dyson}), only evolution in $E$ needs to be taken into account \cite{ZQiuPST2011}, since both $P_{\phi}$ and $\mu$ are conserved for EGAM with $n=0$ and $\omega_G\ll\Omega_{ci}$.

The EGAM equation with the slowly temporal evolving EP ``equilibrium" distribution function obtained from equation (\ref{eq:EP_dyson})  then describes the  evolution of EGAM due to the self-consistent nonlinear interactions of EPs, and exhibits various physics such as wave-particle trapping \cite{TOneilPoF1965,ZQiuPST2011,ABiancalaniJPP2017}, hole and clump pair formation \cite{HBerkPLA1997,HWangPRL2013} and phase-space zonal structure generation and frequency chirping \cite{FZoncaNJP2015,LChenRMP2016}.  This topic  is subject of ongoing research, and an exhaustive analysis is beyond the scope of the present brief review. As illustration and example of nonlinear behavier and particle transport in phase space,  we will briefly introduce the wave-particle trapping in the weak drive limit. We will also qualitatively discuss the secular dynamics in the strong drive limit.

In the weak drive limit,   EGAM   saturation  due to the wave-particle trapping can be demonstrated using test particle approach; with resonant EP orbit only slightly modified.  For simplicity, we consider the $T_e/T_i\ll1$ case, and EGAM is characterized by radial electric field only. Noting that wave-particle energy exchange, is induced by the particle radial acceleration  associated with the radial magnetic drift
$\dot{E}=(e/m)\mathbf{V}_d\cdot\delta \mathbf{E}_r$  and $\mathbf{\dot{R}}=v_{\parallel}\mathbf{b}+\mathbf{V}_d+\delta \mathbf{V}_E$, with $\delta\mathbf{V}_E$ being the $\mathbf{E}\times\mathbf{B}$ drift induced by radial GAM electric field, one then has
\begin{eqnarray}
\dot{v}_{\parallel}=\frac{e}{2mv_{\parallel}}\hat{V}_{dc}\delta E_r\sin\Theta,\nonumber
\end{eqnarray}
where $\Theta=\theta-\omega t+k_G r$ is the phase of resonant particles in the slowly varying wave frame, and $\hat{V}_{dc}=
v^2_{\parallel}/( \Omega R_0)$ is the magnetic curvature drift. Noting that, $\dot{\Theta}=\dot{\theta}-\omega+k_G\dot{r}$, with $\dot{\theta}=\omega_{tr}+\delta V_E/r$, and averaging over fast varying scales, one   obtains
\begin{eqnarray}
\ddot{\Theta}=\frac{e}{2mv_{\parallel}qR_0}\hat{V}_{dc}\delta E_r\sin\Theta.
\end{eqnarray}

This is the typical pendulum equation \cite{TOneilPoF1965,TOneilPoF1968,TOneilPoF1971}, describing the resonant EP being trapped by and exchanging energy with EGAM. When the  wave-particle trapping frequency, $\omega_{B}\equiv\sqrt{e\hat{V}_{dc}\delta E_r/(2m_iv_{\parallel,res}qR_0)}$ is comparable to the EGAM linear growth rate, the mode enters the nonlinear dynamics phase and eventually saturate; as shown by numerical simulations \cite{DZarzosoPoP2012,ABiancalaniJPP2017}.  In this limit, the resonant EP  trajectory is only slightly modified with respect to its equilibrium orbit due to pitch angle scattering, and the drift orbit center position is unchanged; as a result, there is  no EP loss.

In the strong drive limit, however, EP loss may be induced by pitch angle scattering \cite{RNazikianPRL2008}. EGAM self-consistent evolution   can be understood in analogy with the secular fishbone paradigm \cite{RWhitePoF1983,LChenPRL1984,FZoncaNJP2015,GVladNF2013,XWangPRE2012,HZhangPRL2012}. Taking well-circulating EPs as example, the nonlinear evolution of EGAM dominated by nonlinear phase-locking \cite{FZoncaNJP2015} can be qualitatively speculated as follows:  resonant EP parallel velocity, and thus, EP transit frequency decreases as it passes energy to EGAM through transit resonance; and EGAM frequency dominated by EP characteristic frequency decreases consequently. The frequency downward chirped EGAM can keep in phase with EPs losing energy,  leading to nonadiabatic EGAM downward frequency chirping and resonant EP phase space structure secular evolution towards magnetically trapped particle boundary, similar to the ``wave-particle pumping" of fishbones \cite{RWhitePoF1983}.  EPs are lost as they pass the trapped-passing boundary, and become barely trapped particles with unconfined banana orbits, characterized by radial  width comparable with torus minor radius.   This subject is topic of ongoing research, and will be presented in a future publication.


\section{Nonlinear GAM excitation by DWs}
\label{sec:NL_GAM_DW}

The ultimate interest of the fusion community in GAMs is motivated by its  potential interactions with DWs/DAWs and thus, by its positive effect in regulating turbulences and transport \cite{TSHahmPoP1999,ZLinScience1998,LChenPoP2000,FZoncaEPL2008}. This  is achieved via spontaneous excitation of GAM by DWs turbulences, and by scattering of the driving DWs into stable short radial wavelength domain.  The nonlinear excitation of GAM by DWs can be described by a parametric decay instability \cite{RSagdeevbook1969,PKawPoF1969}, where pump  DW resonantly decay into a GAM and another DW. 
GAM nonlinear excitation by DW  has been investigated by analytical theory \cite{NChakrabartiPoP2008,PGuzdarPPCF2008,PGuzdarPoP2009,NChakrabartiPoP2007,JYuJPP2012,JYuPS2010,ZQiuPoP2014,ZQiuNF2014,FZoncaEPL2008}, numerical simulation \cite{RHagerPoP2012,RHagerPRL2012,JLangPoP2007,JLangPoP2008,TDannertPoP2005,RWaltzPoP2008,FLiuPoP2010,XLiaoPoP2016}. The underlying three-wave interactions has also been observed experimentally \cite{GXuPRL2003,TLanPoP2008,WZhongNF2015,AMelnikovNF2017}. In Sec. \ref{sec:NL_GAM_DW}, we will briefly review  these nonlinear wave-wave  interactions in the  gyrokinetic theoretical framework, and emphasize the effects of  kinetic dispersiveness and mode structure associated with realistic geometry and system nonuniformity; which can all affect the nonlinear GAM excitation process qualitatively. As a result, to quantitatively understand and predict fluctuation induced transport,  kinetic treatment and realistic geometry must be properly accounted for.

\subsection{Theoretical model}
\label{sec:NL_GAM_DW_model}

We start with the nonlinear excitation of GAM by DW turbulence. The corresponding  gyrokinetic theory was first presented in Ref. \citenum{FZoncaEPL2008}, while  the detailed derivation was given later in Ref. \citenum{ZQiuPoP2014}. Kinetic treatment is needed here,  since  the nonlinear coupling  increases with increasing $|k_{\perp}\rho_{ti}|$ \cite{FZoncaEPL2008} while the kinetic dispersiveness associated with finite $k_{\perp}\rho_{ti}$ would significantly affect the nonlinear cross-section \cite{MRosenbluthPRL1972,ZQiuPoP2014}. 
The nonlinear equations for the GAM-DW system can   be obtained from the quasineutrality condition, with the nonadiabatic particle response derived from nonlinear gyrokinetic equation \cite{EFriemanPoF1982}. Separating the linear and nonlinear response as  $\delta H\equiv \delta H^L+\delta H^{NL}$, and applying the $\omega\gg\omega_{tr,i},\omega_{d,i}$ assumptions while
solving for the nonlinear ion responses,  one then obtains \cite{FZoncaPoP2004},
\begin{eqnarray}
&&\frac{n_0e^2}{T_i}\left(1+\frac{T_i}{T_e}\right)\delta\phi_k-\langle eJ_k\delta H^L_i\rangle_k+\langle e\delta H^L_e\rangle_k\nonumber\\
&=&-\frac{i}{\omega_k}\left\langle e \Lambda^{\mathbf{k}}_{\mathbf{k'},\mathbf{k''}}  \delta\phi_{k'}\delta H_{e,k''}\right\rangle_k\nonumber\\
&&-\frac{i}{\omega_k}\left\langle e\Lambda^{\mathbf{k}}_{\mathbf{k'},\mathbf{k''}}\left(J_kJ_{k'}-J_{k''}\right)\delta\phi_{k'}\delta H_{i,k''}\right\rangle_k.\label{eq:QN_NL_reduced}
\end{eqnarray}

The first term on the right hand side (RHS) of equation (\ref{eq:QN_NL_reduced})   is formally $O(1/(k^2_{\perp}\rho^2_{ti}))$ larger than the second term from polarization nonlinearity \cite{AHasegawaPoF1978},  for modes with $k_{\perp}\rho_{ti}\ll1$.  However, for the nonlinear GAM equation, the contribution from the first term vanishes due to $\delta H_{d,e}=0$, and the nonlinear GAM equation, then becomes
\begin{eqnarray}
&&\frac{n_0e^2}{T_i}\left(1+\frac{T_i}{T_e}\right)\delta\phi_G-\left\langle eJ_G\delta H^L_{i}\right\rangle_G+\left\langle e\delta H^L_{e}\right\rangle_G \nonumber\\
=&-&\frac{i}{\omega_G}\left\langle e\Lambda^{\mathbf{k}}_{\mathbf{k'},\mathbf{k''}}\left(J_{k'}-J_{k''}\right)\right.\nonumber\\
&&\hspace*{2em}\times\left.\left(\delta\phi_{k'}\delta H_{i,k''}+\delta\phi_{k''}\delta H_{i,k'}\right)\right\rangle_G. \label{eq:NL_para_GAM}
\end{eqnarray}

On the other hand, for nonlinear DW equation, noting  that $\delta H_{d,e}=0$ while $\delta H_{G,e}\neq0$, there is no commutative cancellation in the first term on the RHS of equation (\ref{eq:QN_NL_reduced}), and the DW equation reduces to
\begin{eqnarray}
&&\frac{n_0e^2}{T_i}\left(1+\frac{T_i}{T_e}\right)\delta\phi_k-\langle eJ_k\delta H^L_i\rangle_k\nonumber\\
&=&e\frac{c}{B}\frac{1}{\omega_k}k'_{\theta}\delta\phi_{k'}\frac{\partial\langle\delta H_{e,G}\rangle}{\partial r},\label{eq:NL_para_DW}
\end{eqnarray}
with the selection rule $\mathbf{k}=\mathbf{k}'+\mathbf{k}_G$.

Note that  equations (\ref{eq:NL_para_GAM}) and (\ref{eq:NL_para_DW}) are derived using the $k_{\perp}\rho_{ti}\ll1$ and $1/q\ll1$ expansions, while no assumptions on the mode amplitudes are made except the gyrokinetic ordering \cite{EFriemanPoF1982}. As a result, equations (\ref{eq:NL_para_GAM}) and (\ref{eq:NL_para_DW}) are  general,  and can be applied to study the nonlinear saturation of DWs \cite{ZGuoPRL2009,ZQiuEPS2014,RSinghPoP2014}. In this paper, for the sake of simplicity, we will only review the results obtained for  the ``linear" growing stage of the parametric instability,  with the emphasis on the effect of    system nonuniformities and kinetic dispersiveness on GAM excitation. The nonlinear dynamics of the coupled DW-GAM system including saturation is beyond the scope of this review and,    in fact,  it is still under active investigation.

  \begin{figure}
 \includegraphics[width=3.5in]{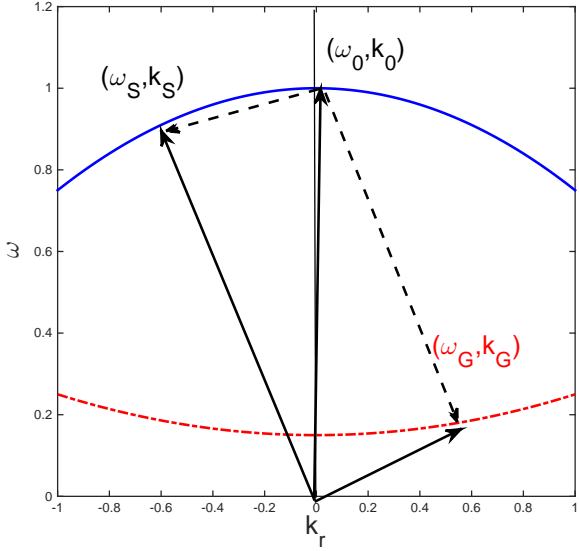}
\caption{Parametric decay of a pump DW into a GAM and a DW lower sideband} \label{fig:parametric_decay}
 \end{figure}

Consistent with the linear growth stage of the parametric instability of interest, through out Sec. \ref{sec:NL_GAM_DW}, we separate the DW into a pump  $\Omega_0(\omega_0,\mathbf{k}_0)$ with finite and fixed amplitude and its lower sideband  $\Omega_S(\omega_S,\mathbf{k}_S)$ with much smaller intensity. Thus, we investigate the resonant decay of the pump DW into a GAM $\Omega_G(\omega_G,\mathbf{k}_G)$ and the lower sideband;
while the feedback of the two daughter waves, i.e.,  $\Omega_G$ and $\Omega_S$, on the pump DW, playing important role in the spectrum evolution and transport, is beyond the scope of this work.  Note that, compared to ZFZF generation by DWs \cite{LChenPoP2000,PGuzdarPoP2001,LChenPRL2004,FZoncaPoP2004}, where nonlinear interactions with  both lower and upper DW sidebands are considered, only the lower DW sideband satisfying the resonant decay condition is considered here, as shown in Fig. \ref{fig:parametric_decay}.   The governing  nonlinear equations can be derived from equations (\ref{eq:NL_para_GAM}) and (\ref{eq:NL_para_DW}),  taking  $\delta\phi_d=\delta\phi_P+\delta\phi_S$,  with   the ballooning mode decomposition for $\delta\phi_d$:
\begin{eqnarray}
\delta\phi_P&=&A_Pe^{-i n\xi-i\omega_Pt}\sum_me^{im\theta}\Phi_0(nq-m)+c.c.,\nonumber\\
\delta\phi_S&=&A_Se^{in\xi-i(\omega_G-\omega_P)t}\sum_me^{-im\theta}\Phi^*_0(nq-m)+c.c.,\nonumber\\
\delta\phi_G&=&A_Ge^{-i\omega_Gt}+c.c.;\nonumber
\end{eqnarray}
and the   eikonal Ansatz   for the radial envelopes; i.e.,
\begin{eqnarray}
A_P&=&e^{i\int k_P dr},\nonumber\\
A_{S}&=&e^{-i\int k_P dr}\left(e^{i\int k_G dr}+c.c.\right),\nonumber\\
A_G&=&e^{i\int k_G dr}+c.c..\nonumber
\end{eqnarray}
Here, $\Phi_0$ accounts for the fine radial scale structure due to finite $k_{\parallel}$ and magnetic shear, with the characteristic radial scale being of  the order of the distance between neighbouring mode rational surfaces, and the normalization  condition $\int ^{\infty}_{-\infty}|\Phi_0|^2 dr=1$ is assumed without loss of generality.
One  then has
\begin{eqnarray}
 D_S \partial_t A_S&=&-\frac{c}{B}k_{\theta,P}k_{r, G}\frac{T_i}{T_e}A^*_PA_G,\label{eq:DWSB_equation_para}\\
 \mathscr{E}_{GAM}\partial_t \delta\phi_G&=&-\alpha_i\frac{c}{2B}k_{\theta,P}k^3_{r,G}\rho^2_{ti} A_SA_P,\label{eq:GAM_equation_para}
\end{eqnarray}
with $\mathscr{E}_{GAM}$ being the linear  GAM dielectric dispersion function \cite{FZoncaEPL2008} defined as
\begin{eqnarray}
\mathscr{E}_{GAM}\equiv 1+T_i/T_e+ T_i \left(\left\langle \delta H_{e}\right\rangle-\left\langle J_0\delta H_{i}\right\rangle  \right)_G/\left( en_0 \delta\phi_G\right),\nonumber
\end{eqnarray}
$\alpha_i=1+\delta P_{\perp}/(en_0\delta\phi_P)$ being an order unity function of local equilibrium parameters defined in Ref. \cite{LChenPoP2000}, and $D_S\equiv D_P(\omega_S, \mathbf{k}_S, r)$ with $D_P$  the linear DW dispersion function  formally defined by
\begin{eqnarray}
D_P\equiv 1+\frac{T_i}{T_e}-\int^{\infty}_{-\infty}\Phi^*_0\left\langle eJ_P\delta H^L_{P,i}\right\rangle dr  /\left(\frac{n_0e^2}{T_i}A_P\right).\nonumber
\end{eqnarray}

For DWs with typically  quadratic dispersiveness, a model dispersion function can be assumed, i.e., $D_P=\omega-\omega_{*0}\exp(-(r-r_0)^2/L^2_*)+C_d\omega_{*0}\rho^2_{ti}k^2_r+iD_I$. $\omega_{*0}$ is the diamagnetic drift frequency at the   gradient peak ($r_0$), and the  Guassian profile indicates a localized  instability drive around gradient peak. We then have
\begin{eqnarray}
D_S&=&i\partial_t+i\gamma_S-\omega_P+\omega_{*0}\left(1-\frac{(r-r_0)^2}{L^2_*}\right)\nonumber\\
&&+C_d\omega_{*0}\rho^2_i\frac{\partial^2}{\partial r^2}.\nonumber
\end{eqnarray}

Defining $\mathscr{E}_G=\partial_r\delta\phi_G/\alpha$, with $\alpha=i(\alpha_i\omega_PT_e/T_i)^{1/2}$, we obtain the following coupled nonlinear equations \cite{LChenVarenna2010}:
\begin{eqnarray}
D_SA_S&=&i\Gamma^*_0\mathscr{E}_G,\label{eq:DWSBequation}\\
D_G\mathscr{E}_G&=&-\Gamma_0\partial_t\partial^2_rA_S,\label{eq:GAMequation}
\end{eqnarray}
in which  $\Gamma_0\equiv(\alpha_iT_i/\omega_PT_e)^{1/2}ck_{\theta,P}A_P/B$ is the normalized pump amplitude, $D_G=(\partial^2_t+\omega^2_G(r)-(G/2)\omega^2_G(r_0)\rho^2_{ti}\partial^2_r)$  with the kinetic dispersiveness term (proportional to $G$) due to FLR/FOW of GAM, and the expression of $G$ can be obtained from equation (\ref{eq:GAM_kinetic_DR_real}) (or equation (31) of Ref. \citenum{ZQiuPPCF2010}; cf. also Sec. \ref{sec:EGAM}).

Equations (\ref{eq:DWSBequation}) and (\ref{eq:GAMequation}) are the coupled nonlinear DW sideband and GAM equations, and   describe the nonlinear parametric generation of these spectral components  by the fixed amplitude pump DW, while the feedback of GAM and DW sideband on the pump DW is neglected due to the $|\delta\phi_G|,  |\delta\phi_S|\ll|\delta\phi_P|$ ordering.  Note that  there are several different radial scales involved in equations (\ref{eq:DWSBequation}) and (\ref{eq:GAMequation}), i.e., the pump DW radial envelope scale $L_P$,  the scale length of diamagnetic drift frequency $L_*$,  and the GAM continuum scale length $L_G$. Note that  we typically have $L_P\sim\sqrt{\rho_{ti}L_*}\ll L_*, L_G$,  and the global DW-GAM problem can then be simplified due to spatial scale separation, with system nonuniformities enter at different spatial scales. The local theory for GAM excitation is presented in Sec. \ref{sec:NL_GAM_local}, while the  role of system nonuniformities is analyzed in Sec. \ref{sec:NL_GAM_global}. The extension of the present theory, largely based on  the $k_{\perp}\rho_{ti}\ll1$ expansion, to short wavelengths and its application to collisionless trapped electron mode with typically $k_{\perp}\rho_{ti}\sim O(1)$  is carried out in Sec. \ref{sec:NL_GAM_CTEM}. Electromagnetic effects are discussed in Sec. \ref{sec:NL_GAM_TAE}.

\subsection{GAM excitation by DWs: local theory}
\label{sec:NL_GAM_local}

\subsubsection{GAM excitation by DWs: parametric   dispersion relation}
\label{sec:NL_GAM_DW_local}

We start from the local limit of the general theory presented in Sec. \ref{sec:NL_GAM_DW_model}, which is  discussed in most publications \cite{NChakrabartiPoP2008,PGuzdarPPCF2008,PGuzdarPoP2009,NChakrabartiPoP2007,FZoncaEPL2008}. Thus,  all the system nonuniformities are neglected, and we focus on the nonlinear interaction strength, defined by the relevant cross-section; i.e., the  coefficients of the nonlinear couplings. Furthermore,  taking $\partial_t=-i\omega+\gamma$ and $\partial_r=i k_r$ in equations (\ref{eq:DWSBequation}) and (\ref{eq:GAMequation}), one then has
\begin{eqnarray}
(\gamma+\gamma_S)(\gamma+\gamma_G)=k^2_G\Gamma^2_{0}, \label{eq:NL_para_DR_local}
\end{eqnarray}
with $\gamma_S$, $\gamma_G$ being the damping rates of DW sideband and GAM, respectively.  In deriving the  above local parametric instability  dispersion relation, the frequency and wavenumber matching conditions for resonant decay illustrated in Fig. \ref{fig:parametric_decay}, are applied, i.e.,
\begin{eqnarray}
\omega-\omega_P+\omega_*-C_d\omega_*\rho^2_{ti}k^2_G&=&0,\nonumber\\
\omega^2-\omega^2_G-G\omega^2_G\rho^2_{ti}k^2_G/2&=&0,\nonumber
\end{eqnarray}
corresponding to energy and momentum conservation in the parametric decay process.

The threshold condition  for GAM  spontaneous excitation is  then given by $k^2_G\Gamma^2_0=\gamma_S\gamma_G$; while,  in the strong drive limit with  the pump DW amplitude well above threshold, the GAM growth rate  is  $\gamma=k_G\Gamma_0$. Note that the nonlinear drive increases with $k_G$, i.e., the generation of  short wavelength KGAM is preferred. This provides   the motivation for the kinetic treatment here, especially when the group velocities of DW sideband and GAM, proportional to $k_G$, are accounted for. This  also motivates  deriving  the short wavelength KGAM dispersion relation, especially the damping rate $\gamma_G$ in Sec. \ref{sec:GAM_kinetic_theory}  that determines the threshold condition for the parameter regime of practical interest.

Before the discussion of global properties of the parametric instability, we would like to briefly discuss the  extensions of the present model,   summarized by governing  equations (\ref{eq:DWSBequation}) and (\ref{eq:GAMequation}) and derived based on  the $k_{\perp}\rho_{ti}\ll1$ expansion for electrostatic DWs, to short wavelengths $k_{\perp}\rho_{ti}\sim O(1)$ and its application to CTEM DW \cite{ZQiuNF2014}. We also generalize our analysis to electromagnetic limit with application to GAM excitation by TAE \cite{ZQiuEPL2013}. These two different cases  are described by governing equations with   forms similar to equations (\ref{eq:DWSBequation}) and (\ref{eq:GAMequation}), despite nonlinear terms have different origin and structure. As a result, the global properties discussed in Sec. \ref{sec:NL_GAM_global} can be, at least  qualitatively, applied to the processes discussed in Sec. \ref{sec:NL_GAM_CTEM} and \ref{sec:NL_GAM_TAE}.

\subsubsection{GAM excitation by short wavelength CTEM}
\label{sec:NL_GAM_CTEM}

The   kinetic  theories of GAM excitation by DWs discussed so far  are derived based on the small argument  expansion of the Bessel functions accounting for FLR effects. This is generally not applicable to   CTEM DW \cite{JAdamPoF1976,WTangNF1978,PCattoPoF1978,CZChengNF1981} with typically $k_{\perp}\rho_{ti}\sim O(1)$.   Another major difference of CTEM with ITG lies in the electron kinetic response, which is also expected to affect   the nonlinear CTEM dynamics, including the excitation of GAM.  The excitation of GAM by CTEM is of interest because  GAM is preferentially excited in the plasma edge, where GAM Landau damping rate is minimized due to its dependence $q$, and where CTEM are also localized due to the fraction of trapped electrons increasing with $r/R_0$.
Numerical simulations using core plasma parameters suggest that GAM excitation is not important for CTEM nonlinear dynamics \cite{DErnstPoP2009,YXiaoPRL2009}, while possible important role of GAMs in regulating CTEM turbulence is observed in simulations using  edge-like parameters \cite{FLiuPoP2010}.  The analytical theory for GAM excitation by CTEM was developed in Ref. \citenum{ZQiuNF2014}, with emphasis on dominant contributions on nonlinear couplings from ions and electrons in different wavelength regimes.

The corresponding nonlinear GAM equation, with an expression similar to equation (\ref{eq:GAM_equation_para}), can be derived as 
\begin{eqnarray}
\mathscr{E}_{GAM} A_G=i\frac{c}{B_0}k_Gk_{\theta}\frac{1}{\omega^2_0}\left(F_1-\alpha_e\right)A_PA_S. \label{eq:NL_GAM_equation_CTEM}
\end{eqnarray}
Here, 
\begin{eqnarray}
F_1=\left\langle J_0J_SJ_GF_0\overline{\frac{\omega_G\omega_{*,i}+(\omega_0-\omega_{*,i}) \omega_{dr}}{\omega_G-\omega_{dr}}}\right\rangle\nonumber
\end{eqnarray}
 is due to ion nonlinearity, and
 \begin{eqnarray}
\alpha_e\equiv -\frac{T_i}{T_e}B_0\int EdEd\Lambda \left|\overline{\overline{\sum_m e^{(i(nq-m)\theta)}}}\right|^2 F_0\omega_{*,e}\oint\frac{d\theta}{v_{\parallel}}\nonumber
\end{eqnarray}
is related to the trapped electron nonlinearity \cite{MRosenbluthPRL1998,FZoncaPoP2004,LChenNF2007}, with $\overline{\overline{(\cdots)}}$ denoting bounce averaging.  In deriving $\alpha_e$, only the contribution of electron temperature  gradient to $\omega_{*,e}$ is considered. 

The CTEM sideband equation, can be derived similarly, 
\begin{eqnarray}
D_S A_S=-i\frac{c}{B_0}k_Gk_{\theta}\frac{1}{\omega^2_0}\left(F_1-\alpha_e\right)A_GA_P^*,\label{CTEM_sideband_equation}
\end{eqnarray}
where $D_S\equiv D_{Ct}(\omega_S,\mathbf{k}_S)$ is the linear dispersion function of CTEM sideband,   and
\begin{eqnarray}
D_{Ct}&\equiv& 1+\frac{T_i}{T_e} - \int^{\infty}_{-\infty}\Phi^*_0\left[\left\langle J_0\delta H^L_{Ct,i}\right\rangle-\left\langle  \delta H^L_{Ct, te}\right\rangle\right]dr\nonumber\\ 
&&\hspace*{4em}\left/\left(\frac{n_0e}{T_i}\int^{\infty}_{-\infty}\Phi^*_0 \delta\phi_{Ct}  dr\right)\right..\nonumber
\end{eqnarray}

Noting $D_S\simeq -i \partial_{\omega_0} D_{Ct, r}\left(\gamma+\gamma_S\right)$, one then obtain  the  following parametric instability dispersion relation
\begin{eqnarray}
(\gamma+\gamma_G)(\gamma+\gamma_S) =\Gamma^2_{D,Ct},\label{eq:NLDR_GAM_CTEM}
\end{eqnarray}
which is  similar to equation (\ref{eq:NL_para_DR_local}) derived in the $|k_{\perp}\rho_{ti}|\ll1$  long wavelength limit. Here, the nonlinear drive due to both ion and trapped electrons is given by
\begin{eqnarray}
\Gamma^2_{D,Ct}\equiv\left(\frac{c}{B_0}k_{\theta}\frac{1}{\omega^2_0}\right)^2\frac{\omega_G}{\rho^2_i\partial D_{Ct,r}/\partial\omega_0}(F_1-\alpha_e)^2|A_0|^2 .\nonumber
\end{eqnarray}
The trapped electron contribution is typically proportional to  $\alpha_e/\omega_*\sim O(\sqrt{\epsilon})$,  while  the ion contribution $F_1$ is sensitive to the perpendicular wavelength $k_{\perp}\rho_{ti}$. Thus,  ions and  trapped electrons contributions dominate in the long and short wavelength limit, respectively. Meanwhile  in the general case with $k_{\perp}\rho_{ti}\sim O(1)$, it can be estimated  that  $F_1$ and $\alpha_e$ are both positive in the simple $\eta_i=\eta_e=0$ limit. The contributions from electrons and ions will, therefore, compete with each other, and thus,  numerical  solution is required for assessing the CTEM parametric decay rate in the general case.  This analysis     is also of    broader interests for  the nonlinear dynamics of kinetic Alfv\'en waves (KAW), e.g., convective cells generation by KAW \cite{FZoncaEPL2015},  nonlinear decay of KAW \cite{LChenEPL2011} and kinetic toroidal Alfv\'en eigenmode \cite{FZoncaPoP1996,RMettPoFB1992,ZQiuPRL2017}.

\subsubsection{GAM excitation by Toroidal Alfv\'en eigenmode}
\label{sec:NL_GAM_TAE}

Alfv\'enic instabilities excited by EPs, e.g., fusion-$\alpha$s, are important for burning plasmas, due to their roles in EP  as well as thermal plasma transport processes, as reviewed in Ref. \citenum{LChenRMP2016}.  Of particular interest is TAE, which exists  in the toroidicity induced SAW continuum gap with minimized excitation threshold \cite{CZChengAP1985,GFuPoFB1989,LChenVarenna1988}. Nonlinear excitation of ZS is one possible channel for Alfv\'enic instability nonlinear saturation \cite{LChenPRL2012,ZQiuPoP2016,ZQiuNF2016,ZQiuNF2017}.  Spontaneous excitation of GAM by TAE was  investigated in Ref. \citenum{ZQiuEPL2013},  demonstrating that the pump TAE is  scattered into a TAE sideband with finite radial envelope due to GAM modulation. The main difference in the electromagnetic TAE case, with respect to the electrostatic DW situation discussed above, is the additional contribution from the nonlinear Maxwell stress term, i.e., the $\mathbf{\delta J}\times\mathbf{\delta B}$ term in momentum equation.  For the SAW related instability  in ideal MHD uniform plasma  limit,  Maxwell stress may cancel    Reynolds stress, yielding the well-known  ``pure Alfv\'enic state" (PAS),  where the Alfv\'enic fluctuation can exist at finite amplitude without significant distortion from nonlinearity      \cite{LChenPoP2013}. The generation of  ZS,  including GAM spontaneous excitation by TAE,  is enabled by the breaking of PAS due to, e.g.,  toroidicity  as an intrinsic nonuniformity of tokamak\cite{LChenPRL2012}.

Nonlinear vorticity equation \cite{LChenJGR1991,LChenNF2001} is needed in addition to the quasi-neutrality condition
\begin{eqnarray}
&&\frac{c^2}{4\pi\omega^2}B\frac{\partial}{\partial l}\frac{k^2_{\perp}}{B}\frac{\partial}{\partial l}\delta\psi_k+\frac{e^2}{T_i}\langle(1-J^2_k)F_0\rangle\delta\phi_k\nonumber\\
&&-\sum\left\langle\frac{q}{\omega}J_k\omega_d\delta H\right\rangle_k=-i\frac{c}{B\omega}\sum_{\mathbf{k}=\mathbf{k}'+\mathbf{k}"}\hat{\mathbf{b}}\cdot\mathbf{k}''\times\mathbf{k}'\nonumber\\
&\times&\left[\left\langle e(J_kJ_{k'}-J_{k''})\left(\delta\phi+\frac{i}{\omega}v_{\parallel}\partial_l\delta\psi\right)_{k'}\delta H_{i,k''}\right\rangle\right.\nonumber\\
&+&\left.\frac{k''^2_{\perp}c^2}{4\pi}\frac{1}{\omega_{k'}\omega_{k''}}\partial_l\delta\psi_{k'}\partial_l\delta\psi_{k''}\right];\label{vorticityequation}
\end{eqnarray}
with the terms on the left hand side being, respectively, field line bending, inertia  and ballooning-interchange terms, and the terms on the RHS being   Reynolds and Maxwell stresses. Furthermore,  $\delta\psi\equiv \omega\delta A_{\parallel}/(ck_{\parallel})$ is defined as an additional variable, and the ideal MHD parallel Ohm's law is recovered if we take $\delta\phi=\delta\psi$.
The particle responses are derived from the   nonlinear gyrokinetic equation, equation (\ref{eq:NLGKE}),  in the $\beta\ll1$ limit, while higher order electro-magnetic component of GAM is neglected. 

Noting $\delta H^L_{T,e}=-(e/T_e)F_0\delta\phi_T$ and $\delta H^L_{T,i}\simeq (e/T_i)F_0J_T\delta\phi_T$, the GAM equation can be derived from the vorticity equation in the form
\begin{eqnarray}
\omega_G\mathscr{E}_{GAM}A_G=-\frac{i}{2}\frac{c}{B}k_{0,\theta} k^3_G\rho^2_{ti}\left(1-\frac{\omega^2_A}{4\omega^2_0}\right)A_SA_0,\label{eq:NLGAM_TAE}
\end{eqnarray}
with $\omega_A\equiv V_A/(qR_0)$, and the two terms in the bracket on the right hand side corresponding to, respectively, the Reynolds and Maxwell stresses.  

Due to the coupling to GAM, the TAE sideband deviation  from ideal MHD can be derived from quasi-neutrality condition
\begin{eqnarray}
\delta\phi=\delta\psi-i\frac{c}{B}k_{0,\theta}k_G\frac{1}{\omega_0}\delta\phi_G\delta\psi^*_0.\label{eq:NL_QN_TAE}
\end{eqnarray}
Substituting into the vorticity equation, one then obtains the nonlinear TAE sideband eigenmode equation
\begin{eqnarray}
k^2_{\perp, S}\epsilon_{T,S}A_S=2i\frac{c}{B}k_{0,\theta}k_G\omega_0k^2_{0,\perp}  A^*_0  A_G,\label{NLTAEsidebandequation}
\end{eqnarray}
where  
$\epsilon_{T,S}=\omega^4_A\Lambda_T(\omega_S)D(\omega_S,k_G)/(\epsilon_0\omega^2_S)$, 
$D(\omega,k_G)=\left(\Lambda_T(\omega)-\delta\hat{W}(\omega,k_G)\right)$
with $\Lambda_T=\sqrt{-\Gamma_-\Gamma_+}$, $\Gamma_{\pm}\equiv(\omega^2/\omega^2_A-1/4)\pm\epsilon_0\omega^2/\omega^2_A$ and $\delta\hat{W}(k_G,\omega)$ playing the role of a normalized potential energy \cite{FZoncaPoFB1993}. Furthermore, $\epsilon_0=2(\epsilon+\Delta')$ with $\Delta'$ being the Shafranov shift in the shifted circular magnetic flux surfaces tokamak case we consider here.   Solutions of $D(\omega,k_G)=0$ are $\omega=\omega_T(k_G)$, with the pump TAE frequency given by $\omega_0=\omega_T(k_G=0)$.

The nonlinear dispersion relation of the parametric instability can be obtained by combining equations (\ref{eq:NLGAM_TAE}) and (\ref{NLTAEsidebandequation})
\begin{eqnarray}
\mathscr{E}_{GAM}\epsilon_{T,S}=\left(\frac{c}{B}k_{0,\theta}k^2_G\rho_{ti}\right)^2\frac{k^2_{0,\perp}}{k^2_{S,\perp}}\frac{\omega_0}{\omega_G}\left(1-\frac{\omega^2_A}{4\omega^2_0}\right) |A_0|^2.\nonumber 
\end{eqnarray}
The nonlinear excitation then replies on the breaking of PAS ($1-\omega^2_A/(4\omega^2_0)\neq0$) by toroidicity. Noting $D(\omega_S,k_G)\simeq-i \partial_{\omega_0} D(\gamma+\gamma_S)$, We thus obtain the   dispersion relation of the parametric decay process
\begin{eqnarray}
(\gamma+\gamma_S)(\gamma+\gamma_G)=\Gamma^2_{D,T};\label{NLDR}
\end{eqnarray}
where the driving term $\Gamma_{D,T}$ is defined as
\begin{eqnarray}
\Gamma^2_{D,T}&\equiv&\left(\frac{c}{B}k_{0,\theta}k_G\right)^2\frac{k^2_{0,\perp}}{k^2_{S,\perp}}\frac{\epsilon_0\omega^3_0}{\omega^4_A\Lambda_T(\omega)}\nonumber\\
&&\times\frac{|A_0|^2}{\partial_{\omega_0} D_0}\left( 1-\frac{\omega^2_A}{4\omega^2_0}\right).
\end{eqnarray}

For typical tokamak parameters, one   has $\omega_0\partial_{\omega_0}D_0>0$. The spontaneous excitation of GAM, thus, requires $\omega^2_0>\omega^2_A/4$, i.e., the pump TAE lies in the upper half of the toroidicity induced gap, which is not the general case. The threshold condition for the parametric instability can be estimated  as $(\delta B_r/B_0)\sim O(10^{-4})$, comparable with other mode-mode coupling channels \cite{TSHahmPRL1995,FZoncaPRL1995,LChenPPCF1998,LChenPRL2012}. Note that, as it was pointed out  in Ref. \citenum{ZQiuNF2017}, ZS excited by weakly ballooning Alfv\'en eigenmodes may have a fine scale radial structure in addition to the well-known meso-scale radial envelope  considered  here, which may further enhance the nonlinear coupling, leading to faster GAM excitation and lower threshold.

\subsection{Nonlinear GAM excitation by DWs: Global theory }
\label{sec:NL_GAM_global}

\subsubsection{Finite DW/GAM dispersiveness: convective amplification and nonlinear GAM group velocity}

When  finite interaction region due to finite pump DW radial envelope is taken into account,  effects of finite GAM and DW sideband group velocities play important roles in the nonlinear dynamics \cite{ZQiuPoP2014}. Neglecting   system nonuniformities due to $\omega_*$ and GAM continuum while retaining finite pump DW radial envelope, i.e., considering a time scale shorter than $L_P/V_c$ with $V_c$ defined later, equations (\ref{eq:DWSBequation}) and (\ref{eq:GAMequation})   become
\begin{eqnarray}
\left(\partial_{\tau}+V_S\partial_{\xi}\right)A_S&=&\Gamma^*_0(\xi)\mathscr{E},\label{eq:NL_para_DWSB_reduced}\\
\left(\partial_{\tau}+V_G\partial_{\xi}\right)\mathscr{E}_G&=& \Gamma_0(\xi)(k^2_r-2ik_r\partial_{\xi})A_S.\label{eq:NL_para_GAM_reduced}
\end{eqnarray}
In deriving the above equations, two temporal and spatial scale expansion, $\partial_t=-i\omega+\partial_{\tau}$ and $\partial_r=ik_r+\partial_{\xi}$, are applied.  Here, $V_S=C_d\omega_{*0}\rho^2_{ti}k_r$ and $V_G=G\omega^2_G(0)\rho^2_{ti}k_r/(2\omega)$ are respectively, the linear group velocities of DW sideband and GAM.   Note that finite dissipation due to $\gamma_S$ and $\gamma_G$ are neglected, as we focus on the global properties of the parametric instability \cite{MRosenbluthPRL1972}.

\begin{figure}
\includegraphics[width=3.5in]{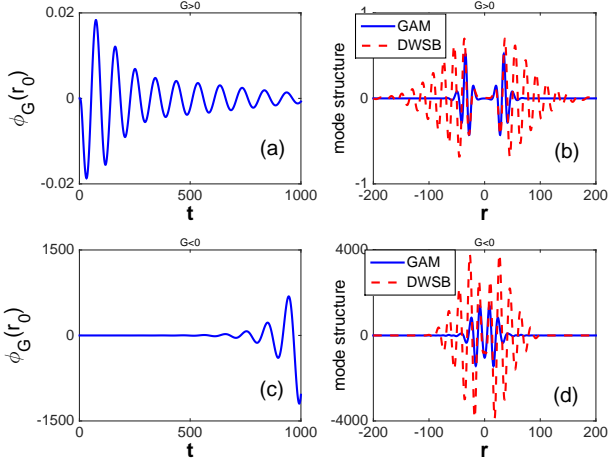}
\caption{ (Reproduced from Fig. 1 of Ref. \citenum{ZQiuPoP2014}.)  Figs. (a) and(c) are respectively the GAM amplitude at $r_0$ v.s. time for $G=\pm1$. Figs. (b) and  (d) are respectively the  snapshot of mode structure at $t=100/\omega_G$ for $G=\pm1$. }\label{fig:NL_para_convectiveabsolute}
\end{figure}

The parametric instability with both daughter waves having a linear group velocity is discussed in Ref. \cite{MRosenbluthPRL1972}. As main result, the instability is a convective amplification process when the  two daughter waves  propagate in the same direction (equivalent to $C_dG>0$ for the case considered here); while   absolute instability exists if the two daughter waves propagate in opposite directions (i.e., $C_dG<0$).  Equations (\ref{eq:NL_para_DWSB_reduced}) and (\ref{eq:NL_para_GAM_reduced}) are solved numerically, with fixed $C_d=1$   and changing the sign of $G$ to explore both cases. The results are shown in Fig. \ref{fig:NL_para_convectiveabsolute}. It is clearly seen that, for $GC_d>0$ (DW sideband and GAM propagate  in the same direction), the parametric instability is a convective amplification process; while for $C_dC_G<0$  it is an absolute instability. Due to the finite pump DW radial width, for $C_dG>0$, the coupled DW sideband and GAM wave packet may propagate out of the unstable region of the parametric instability before they are well developed.  The  value of $C_d$ is typically positive, while the sign of $G$ is investigated carefully in Ref. \citenum{FZoncaEPL2008}. For typical tokamak parameter, we have $G>0$. As a result, the nonlinear excitation of GAM, is typically a convective instability within the present analysis.

The radial propagation of GAM  has been observed in experiments \cite{TIdoPPCF2006,YXuPPCF2011,YHamadaNF2013,DKongNF2013,WZhongNF2015},  investigated in numerical simulations \cite{RHagerPoP2012,ZLiPoP2017,DZarzosoPRL2013},  and  computed  analytically  based on linear  KGAM dispersion relation \cite{FZoncaEPL2008} considering short wavelength structure generation due to the GAM continuous spectrum \cite{ZQiuPST2011,FPalermoEPL2016,ABiancalaniPoP2016}. However, when the experimental data \cite{DKongNF2013} and numerical results \cite{RHagerPoP2012} are compared with KGAM dispersion relation, the obtained coefficient for kinetic dispersiveness  is much bigger than that predicted by linear theory and due to FLR and FOW of ions. The nonlinear velocity of the coupled DW sideband and GAM wave packets discussed above, provide  another interpretation; noting that in experiment \cite{DKongNF2013}, the GAM is driven by ambient turbulence. 

Moving into the wave frame by taking $\zeta=\xi-V_ct$  with $V_c=(V_S+V_G)/2$,  and taking $\mathscr{E}_G=\exp(i\hat{\beta} \zeta)\hat{A}(\zeta,\tau)$ with $\hat{\beta}\equiv k_r\Gamma^2_0/(2V^2_0)$ and $V_0\equiv (V_S-V_G)/2$, the coupled nonlinear equations (\ref{eq:NL_para_DWSB_reduced}) and (\ref{eq:NL_para_GAM_reduced}) can be combined into
\begin{eqnarray}
\left(\partial^2_{\tau}-V^2_0\partial^2_{\zeta}\right)\hat{A}=\left(k^2_r\Gamma^2_0-ik_r\Gamma^2_0\partial_{\zeta}\right)\hat{A}\equiv \hat{\eta}^2\hat{A},
\end{eqnarray}
which can be solved and yields the following unstable solution:
\begin{eqnarray}
\hat{A}&=&\frac{\hat{A}_0}{\sqrt{\pi}\Delta k_0}\int^{\infty}_{-\infty} dk_I \nonumber\\
&&\times \exp\left[-\frac{k^2_I}{\Delta k^2_0}+ik_I\zeta+\sqrt{\hat{\eta}^2-k^2_IV^2_0}\tau\right].
\end{eqnarray}
This is the solution for a typical initial condition $\hat{A}=\hat{A}_0\exp(-\Delta k^2_0 \zeta^2/4)$ at $\tau=0$;  i.e., the parametrically excited GAM has a finite  initial spectrum width $\Delta k_0$.  As the convective damping due to dispersiveness is smaller compared to the temporal growth, i.e.,  $|V_c\partial_{\zeta}|\ll|\partial_{\tau}|$, the time asymptotic solution of GAM electric field is then
\begin{eqnarray}
\mathscr{E}_G=\frac{\hat{A}_0}{\Delta k_0\lambda_{\tau}}\exp\left[\hat{\eta}\tau+i\hat{\beta}(\zeta-V_c\tau)-\frac{(\zeta-V_c\tau)^2}{4\lambda^2_{\tau}}\right].\label{eq:NL_para_GAM_convective}
\end{eqnarray}
From the second term in the exponent, it is clear that $\hat{\beta}$ can be interpreted as the nonlinear modification of the GAM wave vector, while it also affects the GAM frequency through $\hat{\eta}$. $\lambda_{\tau}\equiv \sqrt{1/\Delta k^2_0+V^2_0\tau/(2\hat{\eta})}$  describes  the broadening of the initial GAM pulse during the propagation. 

The solution in equation (\ref{eq:NL_para_GAM_convective})  provides direct information for the interpretation of  experimental observations  \cite{DKongNF2013,DKongNF2017} and/or nonlinear simulations \cite{RHagerPoP2012}. The parametrically excited GAM is characterized by a nonlinear radial wavenumber
\begin{eqnarray}
k_{\footnotesize{NL}}=k_r-i\partial_{\zeta}\ln\mathscr{E}=k_0(1+\Gamma^2_0/(2V^2_0)),\label{eq:NL_wavenumber}
\end{eqnarray}
and a nonlinear frequency
\begin{eqnarray}
\omega_{\footnotesize{NL}}=\omega_0+i\partial_{\tau}\ln\mathscr{E}=\omega_0+\frac{k_0\Gamma^2_0V_c}{2V^2_0}.\label{eq:NL_frequency}
\end{eqnarray}
Both  increase with the pump DW amplitude.  The frequency and wavenumber at vanishing $\Gamma_0$, $(\omega_0,k_0)$,  can be solved from the matching conditions, which can be substituted into equation (\ref{eq:NL_frequency})  and yields,
\begin{eqnarray}
\omega_{\footnotesize{NL}}=\omega_G+\frac{k_0\Gamma^2_0V_c}{2V^2_0}+\frac{G\omega_G\rho^2_{ti}k^2_{\footnotesize{NL}}}{4(1+\Gamma^2_0/(2V^2_0))^2}.\label{eq:NL_para_frequency}
\end{eqnarray}

Note that, $V_c$ and $V_0$ are both proportional to $k_0$, and,  thus, the  frequency increment due to finite amplitude pump DW,  $k_0\Gamma^2_0V_c/(2V^2_0)$, is independent of $k_0$.  The frequency increment, can be expressed as $(e\delta\phi/T)^2(L_n/\rho_{ti})^2$ from our theory, which indicates an order of unity frequency increment for typical parameters. This may explain the existence of the higher frequency branch of the ``dual-GAM" observed in HT-7 tokamak \cite{DKongNF2013}, which has a frequency almost double of the local GAM frequency.

\begin{figure}
\includegraphics[width=3.5in]{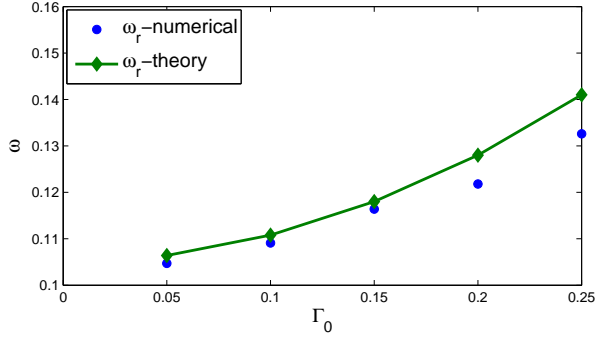}
\caption{(Reproduced from Fig. 3 of Ref. \citenum{ZQiuPoP2015}.)  Dependence of parametrically excited GAM  frequency on pump DW amplitude.} \label{fig:NL_DR_RF}
\end{figure}

\begin{figure}
\includegraphics[width=3.5in]{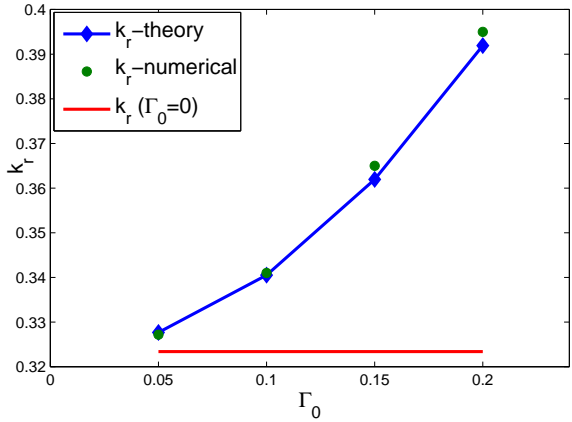}
\caption{(Reproduced from Fig. 1 of Ref. \citenum{ZQiuPoP2015}.) Dependence of parametrically excited GAM  wavenumber on pump DW amplitude.} \label{fig:NL_DR_KR}
\end{figure}

The obtained expressions of the frequency and wavenumber of the parametrically excited GAM, equations (\ref{eq:NL_frequency}) and (\ref{eq:NL_wavenumber}), are compared with the numerical solutions of equations (\ref{eq:NL_para_DWSB_reduced}) and  (\ref{eq:NL_para_GAM_reduced}) shown in Figs. \ref{fig:NL_DR_RF} and \ref{fig:NL_DR_KR}, respectively,  and the analytical solutions fit well with the numerical results.


\begin{figure}
\includegraphics[width=3.5in]{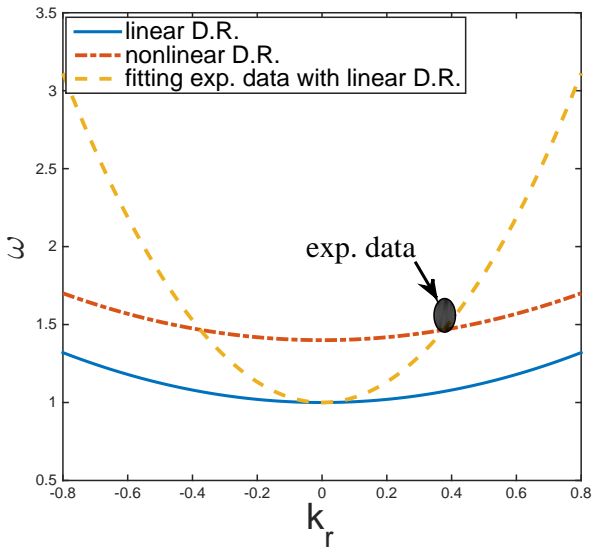}
\caption{Nonlinear dispersion relation of parametrically excited GAM} \label{fig:NL_DR_cartoon}
\end{figure}

The nonlinear dispersion relation of the parametrically excited GAM, $\omega_{\footnotesize{NL}}(k_{\footnotesize{NL}})$, is plotted in Fig. \ref{fig:NL_DR_cartoon}; along with the linear dispersion relation $\omega_0(k_0)$. Note that, the vertical and horizontal axes are, respectively, the ``observed" frequency and wavenumber. The frequency increment due  to finite amplitude pump DW  has a weak dependence on the wavenumber. Thus,  the ``effective" $G$ obtained from experiments \cite{DKongNF2013} or simulations \cite{RHagerPoP2012}  should be smaller than that derived from linear KGAM theory \cite{ZQiuPPCF2009}. However, if only one point is obtained from experiments/simulations  and then fitted with the linear dispersion relation \cite{ZQiuPPCF2009},  overestimation of ``$G$" will be made, as shown by the dashed line. 
From a rough estimation using typical parameters, the misinterpretation may lead to an $O(10^2)$ overestimation of the ``$G$", consistent with that reported in literatures \cite{RHagerPoP2012,DKongNF2013}.

\subsubsection{Nonuniform Plasma: quasi-exponentially growing absolute instability}

Note that, in the above analysis,  we have neglected plasma nonuniformity and, thus,  the analisis is valid for a time scale shorter than $L_P/V_c$.
Next, we consider the longer time scale, and  take the nonuniformity of $\omega_*(r)$  into account, while we neglect  the contribution of GAM continuum in order to illustrate the effects of nonuniform $\omega_*(r)$. Equations (\ref{eq:DWSBequation}) and (\ref{eq:GAMequation})   are solved numerically, and the result  shows that  outward propagating coupled DW sideband and GAM wave packets  are reflected at the DW turning points    due to  $\omega_*(r)$ nonuniformity, and are amplified as they propagate through their original position $r_0$ again. The convective instability, as a result, becomes a quasi-exponentially growing absolute instability.

\begin{figure}
\includegraphics[width=3.5in]{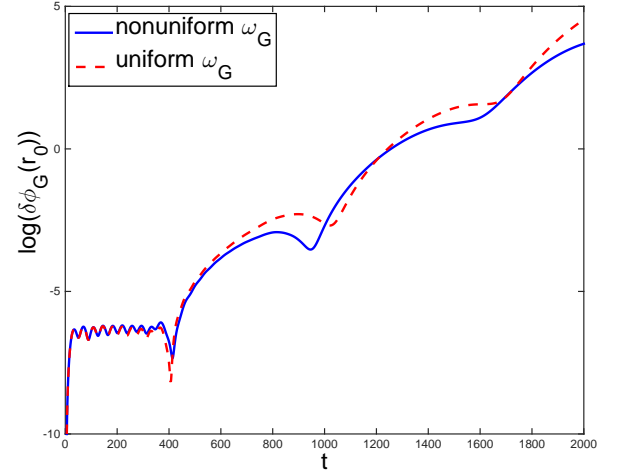}
\caption{(Reproduced from Fig. 6 of Ref. \citenum{ZQiuPoP2014}.)  mode structures at $t=500$} \label{fig:GAMamplitudeNuDWNuP}
\end{figure}

In the strong drive limit with $|\omega^2-\omega^2_G|\simeq|2\gamma \omega_G|\gg|G\omega^2_G\rho^2_ik^2_r/2|$,  the KGAM kinetic dispersiveness term can be ignored, and the coupled  equations can be combined to yield the nonlinear DW sideband eigenmode equation in  Fourier-$k_r$ space \cite{ZQiuPoP2012,RWhiteNF1974}
\begin{eqnarray}
&&\left(\frac{\omega_*}{L^2_*}\frac{\partial^2}{\partial k^2_r}+\omega-\omega_P+\omega_*-C_d\omega_*\rho^2_{ti}k^2_r\right.\nonumber\\
&&\hspace*{8em}+ \left.\frac{\omega k^2_r\Gamma^2_0}{\omega^2-\omega^2_G}\right)A_S=0. \label{eq:NLDWreduce}
\end{eqnarray}
The linear DW eigenmode equation   can be recovered if one ignores the nonlinear term (the term proportional to $\Gamma^2_0$) in equation (\ref{eq:NLDWreduce}), and it  can be solved to yield the finite   extent of the pump DW in  $k_r$ space, and, equivalently, the localization in real space with a typical scale length $\propto \sqrt{L_*\rho_{ti}}$. Including the nonlinear term, equation (\ref{eq:NLDWreduce})  yields the following nonlinear dispersion relation
\begin{eqnarray}
\frac{L^2_*}{\omega_*}(\omega-\omega_P+\omega_*)\tilde{\beta}^2=2l+1,\ \  l=0, 1, 2, 3\cdots\label{eigenmodeofreduced}
\end{eqnarray}
with $\tilde{\beta}$ given by
\begin{eqnarray}
\tilde{\beta}^4\frac{L^2_*}{\omega_*}\left(C_d\omega_*\rho^2_i+\frac{\omega\Gamma^2_0}{\omega^2_G-\omega^2}\right)=1.\nonumber
\end{eqnarray}

The eigenmode structure of DW sideband in Fourier space is given by
\begin{eqnarray}
A_S\propto\exp{\left(-\frac{k^2_r}{2\tilde{\beta}^2}\right)},
\end{eqnarray}
with a radial extent of $|\tilde{\beta}|^{-1}$ \cite{TIdoPPCF2006,KZhaoPPCF2010,GConwayPPCF2008,DKongNF2013}.  This explains the localization of GAM by ``density pedestal"   reported  in Ref. \cite{GConwayPPCF2008}, where GAM can only be observed in the density gradient region where density gradient is sharp (i.e., $L_*$  small compared to the plasma minor radius); whereas GAM can be observed well into the plasma when the pedestal weakens.

\begin{figure}
\includegraphics[width=3.8in]{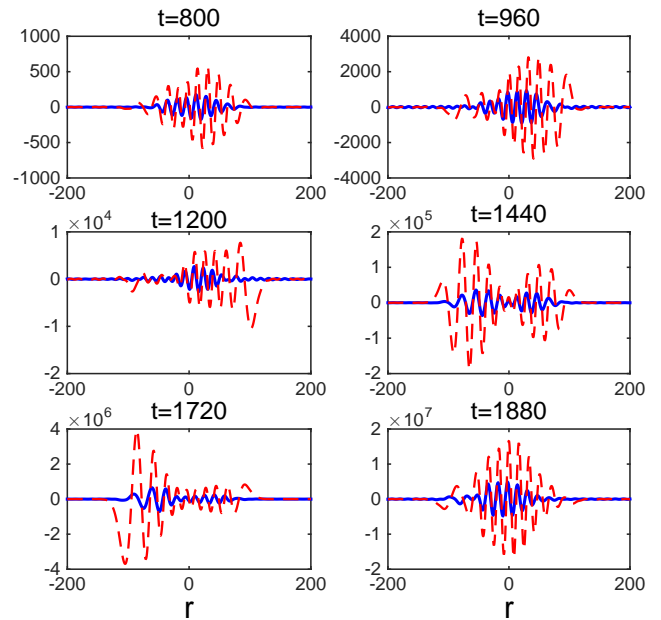}
\caption{(Reproduced from Fig. 5 of Ref. \citenum{ZQiuPoP2014}.)  Snapshots of mode structure at   different times}\label{fig:timesequence}
\end{figure}

Finally, with all the nonuniformities  self-consistently included,   the coupled nonlinear equations  (\ref{eq:DWSBequation}) and (\ref{eq:GAMequation}), are solved numerically. The time histories of GAM amplitude at $r=r_0$ is shown in Fig. \ref{fig:GAMamplitudeNuDWNuP}, in which the solid curve corresponds to the nonuniform GAM frequency case, while the dashed line  illustrates the uniform GAM frequency case for comparison.  One notes  that the two cases are qualitatively similar, i.e., the nonuniformity of $\omega_*(r)$ is the dominant effect on the longer time scale, which renders the initially convective parametric instability into a quasi-exponentially growing absolute instability on a longer time scale. On the other hand, GAM continuum plays a relatively minor role.  Due to the frequency mismatch induced by spatially varying $\omega_G(r)$, the case with nonuniform $\omega_G(r)$ has a slightly different growth rate. The mode structures of coupled DW sideband and GAM
at six different times are shown in Fig. \ref{fig:timesequence}. One may see that,
due to the nonuniformity induced by GAM continuum, the
mode structures propagating in opposite directions are not symmetric.
The wave packet initially propagating outward has a
larger $k_r$ and, thus, larger growth rate and group velocity.
Consequently, one may observe that it also has a larger amplitude;
then, it is reflected at the turning point induced by $\omega_*$ nonuniformity, and propagates inward, completing a full ``bouncing" 
period of wave packets radially trapped by nonuniform $\omega_*$.

Note that, although extensively studied in the past two decades, most publications on the nonlinear interactions of GAM and DW turbulence focuses on the ``linear growth stage" of the parametric instability, i.e., considering     a fixed amplitude DW decays into a GAM and a DW sideband, while the feedbacks of DW sideband and GAM on the pump DW  are neglected.  As a result, the  theories cannot be applied to the nonlinear dynamics of DWs mediated by GAMs, e.g., at saturation.  An attempt is made in Ref. \cite{FZoncaEPL2008}, where the feedbacks of the DW sideband and GAM to the linearly unstable DW pump are considered. The derived driven-dissipative system based on three-wave couplings then exhibits limit-cycle behaviors, period-doubling and route to chaos as possible indication of the existence of strange attractors \cite{FZoncaPoP2004},  which can be applied to interpret experimental observations such as ``predator-prey" behaviors of GAM and DW intensity. However, in the truly nonlinear stage, the strongly modulated DW can no longer be separated as a    pump and a sideband. The two   field model for DW-GAM system,  described by equations  (\ref{eq:NL_para_GAM}) and (\ref{eq:NL_para_DW}),  including full radial wavenumber spectrum should be used, as in the nonlinear dynamics  of the coupled DW-ZFZF system \cite{ZGuoPRL2009}.

\section{Nonlinear self-coupling of GAM/EGAM}
\label{sec:self_NLty}

Nonlinear self-couplings of GAMs were observed in experiments \cite{YNagashimaPPCF2007,RNazikianPrivate2009,LHorvathNF2016}, in the form of   perturbations at GAM second harmonic frequency, and considered to be important  for DW nonlinear dynamics as an additional channel for saturating   GAMs \cite{HZhangNF2009}.  In GTC \cite{ZLinScience1998}  simulations with a finite amplitude GAM as initial condition, scalar potential generation  at GAM second harmonic frequency was observed in the absence of  parallel nonlinearity. However,    GAM  second harmonic generation was suppressed   when parallel nonlinearity was turned on. Analytical theory based on phase  space volume conserving  gyrokinetic equation \cite{ABrizardRMP2007,TSHahmPoF1988,ABrizardPoP1995} explained these simulation results with   the exact cancellation of parallel and perpendicular nonlinearity to the leading order \cite{HZhangNF2009}. No GAM second harmonic scalar potential generation is also obtained from fluid theory, with emphasis on the associated second harmonic  density perturbation  \cite{GFuJPP2011}.  

Even if not emphasized explicitely, the simulations in Ref. \cite{HZhangNF2009}  also show finite ZFZF scalar potential generation by GAM. This process is not affected by the cancellation of  parallel and perpendicular nonlinearities. The analysis based on gyrokinetic theory \cite{LChenEPL2014} shows that  finite ZFZF generation is due to thermal ion FOW effects, so it is a purely ``neoclassical" effect with contribution from toroidal geometry. It is also shown that  there is no  modulation of GAM by ZFZF.  

The GAM second harmonic and ZFZF generation discussed above may have direct impact on the nonlinear dynamics of DW turbulences discussed in Sec. \ref{sec:NL_GAM_DW},  because of the effect of ZFZF on regulating DWs \cite{ZLinScience1998,TSHahmPoP1999,LChenPoP2000,PGuzdarPoP2001,LChenPRL2004,FZoncaPoP2004,PDiamondPPCF2005}. Generation of GAM second harmonic,  which is  not a normal mode of the system, will induce additional dissipation for GAMs. Meanwhile, ZFZF generation by GAM corresponds to direct power transfer from GAM to ZFZS. Both processes will affect the branching ratio of GAM and ZFZF generation by DWs, and, thus, the nonlinear dynamics of DWs.

To understand the GAM second harmonic scalar potential generation, it is shown in Ref. \cite{GFuJPP2011} that  the contribution from resonant EPs will induce EGAM second harmonic scalar potential. Therein, a perturbative model in the small EP drift orbit limit is analyzed  for the simplicity of discussion.  The general theory of  second harmonic and ZFZF generation by EGAM  is given in Ref. \citenum{ZQiuPoP2017}, which can be applied for arbitrary wavelengths.

In the following, the analysis of  Ref. \cite{HZhangNF2009} will be briefly reviewed, with emphasis on the conditions for the cancelation of parallel and perpendicular nonlinearities. The other self coupling channels, investigated in Refs. \cite{GFuJPP2011,LChenEPL2014,ZQiuPoP2017},  will be discussed based on the result of Ref. \citenum{HZhangNF2009}.

\subsection{GAM second harmonic generation}
\label{sec:GAM_2nd_harmonic}

For GAM second harmonic generation by self beating of GAM with $k_{\parallel}=0$, the nonlinear gyrokinetic equation  in the phase-space volume  conserving form \cite{TSHahmPoF1988,ABrizardPoP1995} can be written as:
\begin{eqnarray}
\left(\partial_t+v_{\parallel}\partial_l+i\omega_d\right)\delta F_{II}&=&-\hat{\mathbf{b}}\times\nabla J_G\delta \phi_G\cdot\nabla\delta F_G/B\nonumber\\
& &- \delta \dot{v}_{\parallel}\partial_{v_{\parallel}}\delta F_G,\label{eq:GK_2nd}
\end{eqnarray}
with the first term on the right hand side being the usual perpendicular convective nonlinearity, and the second term being the parallel nonlinearity, which is usually neglected in the   gyrokinetic equation (\ref{eq:NLGKE}). In fact, the latter is typically of higher order when compared with the  perpendicular nonlinearity. Here, $\delta\dot{v}_{\parallel}\equiv -e\mathbf{\hat{b}}\cdot\nabla J_G\delta\phi_G/m-\nabla\times(v_{\parallel}\hat{\mathbf{b}})\cdot\nabla J_G\delta\phi_G/B$, and the subscript ``II" is used for second harmonic.  The GAM second harmonic dispersion relation can be obtained from quasi-neutrality condition, and one has
\begin{eqnarray}
-\frac{e}{m}n_0k^2_{II}\frac{1}{\Omega^2_i}\left(1-\frac{\omega^2_G}{\omega^2_{II}}\right)\overline{\delta\phi}_{II}+\langle \overline{\delta F_{II}}\rangle=0,
\end{eqnarray}
with the second harmonic perturbation derived as
\begin{eqnarray}
\delta F_{II}&=&-\frac{k_G\overline{\delta\phi}_G\hat{F}}{\omega_{II}B_0}\left[\frac{\cos\theta}{r(1-\epsilon\cos\theta)}-\frac{\sin^2\theta}{R_0}\right]. \label{eq:2nd_response}
\end{eqnarray}
Here, $\hat{F}\equiv\delta F_G/\sin\theta\simeq (e/T_i)(\hat{\omega}_{dr}/\omega)F_0J_G\overline{\delta\phi}_G$, and the two terms in equation (\ref{eq:2nd_response}) are respectively,  the perpendicular and parallel nonlinearity contribution.  The perpendicular  nonlinearity is formally $O(\epsilon^{-1})$ larger, as expected, giving  the dominant ``up-down symmetric"  ($\propto\cos\theta$)  second harmonic density perturbation   \cite{RNazikianPrivate2009,GFuJPP2011,HZhangNF2009}.   
However, since the GAM second harmonic dispersion relation is  derived from the surface averaged quasi-neutrality condition, the dominant  perpendicular nonlinearity  proportional to $\cos\theta$, can only have a contribution via the toroidicity term in its denominator, as explicitly given in equation (\ref{eq:2nd_response}). 
As a result, the contribution from parallel   and perpendicular nonlinearity cancels exactly upon taking the flux surface average. Thus, there is no GAM second harmonic scalar potential generation up to the order  of parallel nonlinearity.  

Note that, in the case discussed here, the perpendicular nonlinearity, which is larger by $O(\epsilon^{-1})$, contributes to scalar potential generation through toroidal coupling, making the  contribution $O(\epsilon)$ smaller, and cancels exactly with parallel nonlinearity. Other processes  are then required to have a non-vanishing density perturbation, after surface averaging,  producing finite self coupling of GAMs and generation of GAM second harmonic and/or ZFZF on a time scale shorter than the parallel nonlinearity characteristic time. There are two mechanisms that have been suggested in the literature as possible candidates. One is the   coupling through thermal ion FOW effects,  proposed in Ref. \citenum{LChenEPL2014} for $k_r\rho_{d,i}>\epsilon$; and another one is via EP FOW effects, which are  of larger EP drift orbits but the tradeoff of smaller EP concentration \cite{GFuJPP2011,ZQiuPoP2017}.  For the sake of completeness, we also note that  the symmetry breaking induced by finite amplitude DWs  in GAM second harmonic and ZFZF generation  has been investigated  in the literature \cite{MSasakiPoP2009,MSasakiPPCF2009}, and is related to the long time scale evolution of the coupled DW-GAM system \cite{ZQiuEPS2014,RSinghPoP2014,ZGuoPRL2009}. A detailed discussion of these processes is beyond the scope of the present brief review.

\subsection{ZFZF generation by GAM}

ZFZF generation   is observed in the above mentioned GTC simulations, with or without inclusion of parallel nonlinearity  \cite{HZhangNF2009}. This suggests that  other  mechanisms,   stronger than  toroidal coupling discussed in Sec. \ref{sec:GAM_2nd_harmonic}, may be responsible for the ZFZF generation. Motivated by this evidence, it was shown \cite{LChenEPL2014} that s thermal ion FOW effects   may generate ZFZF scalar potential  for $k_G\rho_{d,ti}>\epsilon$,  with the contribution from perpendicular nonlinearity being significantly larger than parallel nonlinearity after surface averaging.  The nonlinear gyrokinetic equation for ZFZF generation by self beating of GAM, can be written as
\begin{eqnarray}
\left(\partial_t+v_{\parallel}\partial_l\right)\delta H^{NL}_{dZ}=-e^{-ik_Z\rho_d}\frac{1}{B}\hat{\mathbf{b}}\times\nabla \delta \phi_G\cdot\nabla\delta H_G,
\end{eqnarray}
with  $\delta H^{NL}_{dZ}=e^{ik_Z\rho_d}\delta H^{NL}_{Z}$ being the drift orbit center distribution function, $e^{ik_Z\rho_d}$ representing the operator for drift orbit center transformation and $\rho_d=\hat{v}_{d}\cos\theta/\omega_{tr}\equiv\hat{\rho}_d\cos\theta$ being  the drift orbit width defined below equation (\ref{eq:GAM_response_general}).

For ZFZF, with $\omega_Z\ll |v_{\parallel}\partial_l|$, one has $\delta H^{NL}_{dZ}=\overline{\delta H^{NL}_{dZ}}+\widetilde{\delta H^{NL}_{dZ}}\simeq \overline{\delta H^{NL}_{dZ}}$. Therefore,
\begin{eqnarray}
&&\partial_t\overline{\delta H^{NL}_{dZ}}
=- \frac{c}{B_0} \overline{e^{-ik_Z\rho_d}\sum_{\mathbf{k}}\hat{\mathbf{b}}\times\nabla\delta\phi_G\cdot\nabla\delta H_G}.\label{eq:NLZFZF}
\end{eqnarray}

Noting that $\delta\phi_G=\overline{\delta\phi}_G+\delta\phi_{G,1}\sin\theta$, $\rho_d\propto\cos\theta$, $\omega_d\propto\sin\theta$,  assuming $|k_Z \rho_d|\ll1$, using the expression of $\delta H_G$ derived in Sec. \ref{sec:GAM_kinetic_theory} and noting $\omega_G=\omega_0+i\partial_t$,  we then have, after some algebra \cite{LChenEPL2014}
\begin{eqnarray}
\overline{\delta H^{NL}_{dZ}}=-\frac{e}{T_i}F_0\frac{1}{\omega^2_0}\frac{c}{B_0}\hat{v}_{d}\overline{\rho_d\cos\theta}\frac{\partial}{\partial r}\left(\frac{|\delta E_{G,r}|^2}{r}\right).\label{eq:NLZFZFresponse}
\end{eqnarray}
 It is worth  mentioning that    the dominant contribution comes from coupling due to finite drift-orbit width effect; that is,   a neoclassical effect.  Substituting the nonlinear particle response, equation (\ref{eq:NLZFZFresponse}), into the quasi-neutrality condition, we obtain the following nonlinear equation describing nonlinear excitation of ZFZF by a finite amplitude GAM
\begin{eqnarray}
\chi_{Z}\overline{\delta\phi_Z}=-\frac{c}{B_0}\frac{1}{\omega^2_G}\frac{\partial}{\partial r}\left[\left\langle \hat{v}_{d}\overline{\cos\theta \rho_{d}}F_0\right\rangle\frac{|\overline{\delta E_{G}}|^2}{r}\right];
\end{eqnarray}
where, $\chi_Z$ is the well-known neoclassical polarization  \cite{MRosenbluthPRL1998}
\begin{eqnarray}
\chi_Z\overline{\delta\phi_Z}\equiv\left(1-\left\langle \frac{F_0}{n_i}J^2_Z\left|\overline{e^{ik_Z \rho_{d}}}\right|^2\right\rangle\right)\overline{\delta\phi_Z}.\nonumber
\end{eqnarray}
On the other hand, there is no modulation of GAM by ZFZF up to the order of parallel nonlinearity, which is beyond the time scale of interest. Thus, the nonlinear generation of ZFZF by GAM observed in Ref. \citenum{HZhangNF2009} is a forced driven process,  which is, again, underlying the $\omega_G=\omega_{Gr}+i\partial_t$ condition used for  deriving the non-vanishing ion response of ZFZF in eqution (\ref{eq:NLZFZFresponse}).

\subsection{Second harmonic generation by EGAM}

To understand the finite GAM second harmonic scalar potential generation, the effect of EPs was proposed and analyzed in Ref. \citenum{GFuJPP2011}, where resonant EP contribution was treated in the small EP drift orbit limit.  The analysis  is  then extended to arbitrary wavelengths  in Ref. \citenum{ZQiuPoP2017}, for the GAM second harmonic and  ZFZF generation. The basic ideas of Refs. \citenum{GFuJPP2011,ZQiuPoP2017} are consistent with those of Ref. \citenum{LChenEPL2014},  i.e., taking the coupling due to EP FOW into account (noting again $\rho_{d,h}\propto\cos\theta$). In particular,  EPs are characterized by larger drift orbits than thermal ions \cite{LChenEPL2014}; however, EPs have much smaller   density. 

Here,  we will briefly review the approach of  Ref. \citenum{ZQiuPoP2017} using the same gyrokinetic theoretical framework  consistent with the rest of the current review, although the  original analysis is proposed in Ref. \citenum{GFuJPP2011} for  the EGAM second harmonic generation. Generation of ZFZF by EGAM is also investigated in Ref. \cite{ZQiuPoP2017} and  can be derived following the same approach. Again, only   processes   faster than parallel nonlinearity are of interest here.  
Substituting the EP response from equation (\ref{eq:GAM_response_general}) into equation for nonlinear EP drift orbit center distribution function, and considering small but finite $T_e/T_i$, we then obtain the following general expression of the nonlinear EP response to the EGAM second harmonic
\begin{eqnarray}
\delta H^{NL}_{II,h}&=&ik_r\frac{c}{B_0}\partial_EF_{0,h}\sum_{\eta,\xi,p,l}\frac{p+l}{\omega_{II}-(p+\xi+l)\omega_{tr}}\nonumber\\
&\times&i^{\eta+p-\xi-l}e^{i(\eta+p+\xi+l)\theta}J_{\eta}(\hat{\Lambda}_{II,h})J_{\xi}(\hat{\Lambda}_{II,h})\nonumber\\
&\times& J_l(\hat{\Lambda}_h)J_p(\hat{\Lambda}_h)\frac{\omega}{\omega-l\omega_{tr}}\frac{\overline{\delta\phi}_G\overline{\delta\phi}_G}{r}.\label{eq:2nd_EP_general}
\end{eqnarray}
Here, $\hat{\Lambda}_h=k_r\hat{\rho}_{d,h}$ and $\hat{\Lambda}_{II,h}=k_{r,II}\hat{\rho}_{d,h}=2\hat{\Lambda}_h$. Substituting equation (\ref{eq:2nd_EP_general}) into the surface averaged quasi-neutrality condition, we obtain the equation for EGAM second harmonic generation:
\begin{eqnarray}
\hat{b}_{II}\mathscr{E}_{EGAM}(\omega_{II})\frac{en_0}{T_i}\overline{\delta\phi}_{II}=-\left\langle\overline{\delta H^{NL}_{II,h}} \right\rangle,\label{eq:2nd_DR}
\end{eqnarray}
where $\hat{b}_{II}\equiv k^2_{r,II}\rho^2_{L,h}/2$, and $\mathscr{E}_{EGAM}(\omega_{II})$ is the   linear  EGAM dielectric function  at $\omega=\omega_{II}$, with nonadiabatic EP  response given by equation (\ref{eq:GAM_response_general}).  The general dispersion relation obtained from equation (\ref{eq:2nd_DR}) will recover that of Ref. \citenum{GFuJPP2011} in the proper limit, i.e., with $|\Lambda_h|\ll1$ and only resonant EP contributions taken into account. 

For EGAM   with a typically global mode structure, i.e., $|\Lambda_h|\ll1$, the dominant  contribution is obtained for small $|\eta|+|\xi|+|p|+|l|$. Also, $l=\pm1$  can be assumed the strongest linear EGAM drive, and $p+l\neq0$  is required for non-vanishing nonlinear EP response to EGAM second harmonic. With these selection rules in mind,  and  noting that $\omega_{II}=2\omega$ and $\hat{\Lambda}_{II,h}=2\hat{\Lambda}_h$,  one then has
\begin{eqnarray}
\overline{\delta H^{NL}_{II,h}}\simeq i\frac{c}{B_0} \frac{\partial F_{0,h}}{\partial E}\frac{3k_r \omega\hat{\omega}^2_{dr}}{(\omega^2-\omega^2_{tr})(\omega^2_{II}-\omega^2_{tr})}\frac{\overline{\delta\phi}_G\overline{\delta\phi}_G}{r}.\label{eq:2nd_EP_reduced}
\end{eqnarray}

We note that  equation (\ref{eq:2nd_EP_reduced}) is equivalent to equation (51) of Ref. \cite{GFuJPP2011}. Substituting equation (\ref{eq:2nd_EP_reduced}) into the quasi-neutrality condition for EGAM second harmonic, we then obtain:
\begin{eqnarray}
&&\hat{b}_{II} \mathscr{E}_{EGAM}(\omega_{II})\overline{\delta\phi}_{II}\nonumber\\
&=&- \frac{ik_rT_i}{n_0m\Omega_i r}\left\langle   \frac{3\omega\hat{\omega}^2_{dr}\partial_E F_{0,h}}{(\omega^2-\omega^2_{tr})(\omega^2_{II}-\omega^2_{tr})}\right\rangle \overline{\delta\phi}^2_G, \label{eq:2nd_DR_reduced}
\end{eqnarray}
with $\mathscr{E}_{EGAM}(\omega_{II})$ obtained from the proper limit of the linear EGAM second harmonic dispersion relation for  small magnetic drift orbits and  only primary transit resonance accounted for in the nonadiabatic EP response.  Note that,  in   equation (52) of Ref. \cite{GFuJPP2011}, $\omega_{EGAM}$ should  also be a function of $\omega_{II}$ ($\omega_2$ using the notation of Ref. \cite{GFuJPP2011}). Equation (\ref{eq:2nd_DR_reduced})  or,  more precisely, equation (\ref{eq:2nd_DR})  can then be applied to explain experimental observations/simulation results on EGAM second harmonic generation, by directly substituting parameters into the nonlinear dispersion relation  along with both the amplitude and radial mode structure of the primary mode.

\section{Unified theoretical framework of GAM/EGAM}
\label{sec:framework}

The   physics processes discussed above  can be synthetically included into the following ``unified theoretical framework" of GAM/EGAM \cite{ZQiuEPS2014};  including self-consistent generation of GAM by DW turbulences and/or EPs, modulation of DW by GAM/EGAM, and self-consistent evolution of EP equilibrium distribution function due to nonlinear interactions with GAMs. The corresponding equations are
\begin{eqnarray}
\omega_dD_d  A_d&=&\frac{c}{B_0}\frac{T_i}{T_e} k_{\theta} A_d\partial_r  A_G, \label{eq:dw}\\
\omega_G \mathscr{E}_{EGAM}  \partial_rA_G&=& -\frac{\alpha_i c}{ B_0} k_{\theta} \left(A_d\partial^2_r A^*_d-c.c.\right).\label{eq:gam}
\end{eqnarray}
Here, $\mathscr{E}_{EGAM}$ is  the EGAM dispersion relation obtained from equation (\ref{eq:QN_EGAM})   
\begin{eqnarray}
\mathscr{E}_{EGAM}\equiv-1+\frac{\omega^2_G(r)}{\omega^2}-\frac{G}{2}\rho^2_{ti}\partial^2_r+\frac{\overline{\delta n}_h m_i\Omega^2_i}{en_0k^2_r\overline{\delta\phi}_G},\nonumber
\end{eqnarray}
with the perturbed EP density $\delta n_h$ give by equation (\ref{eq:EP_density_perturbation}), where the slowly varying  EP ``equilibrium" distribution function  $F_{0,h}$ due to emission and reabsorption of EGAM is the solution of  the Dyson equation (\ref{eq:EP_dyson}). Thus, this ``unified theoretical framework", based on  equations  (\ref{eq:EP_dyson}),  (\ref{eq:dw}) and (\ref{eq:gam}),  fully describes GAM related physics   in realistic geometries;  including linear physics of GAM/EGAM, nonlinear dynamics of EGAM and nonlinear dynamics of the coupled   GAM/EGAM-DW  system. Note that EP interaction with DWs is typically weak \cite{WZhangPRL2008,ZFengPoP2013}.   We incidentally note that the $\mathscr{E}_{EGAM}$ expression defined here \cite{ZQiuPPCF2010}, besides the EP contribution in $\delta n_h$, has a coefficient $k^2_r\rho^2_{ti}/2$ compared to $\mathscr{E}_{GAM}$ \cite{ZQiuEPL2013,ZQiuPoP2014,ZQiuNF2014} used in Sec. \ref{sec:NL_GAM_DW}  due to the different notations used in original papers. 
Note also that    equations  (\ref{eq:dw}) and (\ref{eq:gam}) are derived based on $k_{\perp}\rho_{ti}\ll1$ expansion;  while no separations of DW into  pump and   sidebands is assumed. As a result, neglecting EP effects, the DW-GAM system described by the two field model, equations  (\ref{eq:dw}) and (\ref{eq:gam}), can be applied to understand the fully nonlinear evolution of DWs, including turbulence spreading and saturation  due to the envelope modulation by GAMs \cite{ZGuoPRL2009}.  Meanwhile, when $A_d$ is separated into a fixed amplitude pump DW and its sideband due to GAM modulation (distortion of parallel mode structure is not significant in nonlinear processes with $\tau_{NL}\gg\omega^{-1}$),  equations  (\ref{eq:DWSBequation}) and (\ref{eq:GAMequation}) are recovered, as shown in Ref. \citenum{ZQiuEPS2014}.

Linear excitation and nonlinear evolution of EGAM, on the other hand, can be described by equations  (\ref{eq:EP_dyson})  and (\ref{eq:gam})  in the absence of DWs.  If, for example, the  equilibrium EP distribution function is used, equation (\ref{eq:gam}) then describes the linear EGAM excitation, as  discussed in Sec. \ref{sec:EGAM_local} and \ref{sec:EGAM_global}. When the slow  EP distribution function evolution on transport time scale due to emission and re-absorption of EGAM is taken into account,  equations   (\ref{eq:EP_dyson}) and  (\ref{eq:gam})   could then provide the self-consistent   EGAM nonlinear dynamics qualitatively discussed in Sec. \ref{sec:EGAM_saturation}.    Thus, the ``unified theoretical framework  of GAM/EGAM" includes all the physics presented in this review. It   also provides the outlooks for possible future  research  on the dynamics evolution of the  fully nonlinear system.

\section{Conclusions and Discussions}
\label{sec:conclusion}

In this paper, the recent theoretical understandings of GAMs   are briefly reviewed; including the linear dispersion properties, resonant excitation by EPs, nonlinear excitation by DWs/DAWs, and the nonlinear self-coupling of GAM/EGAM. The emphasis is  on the effects of system nonuniformities,   the  requirements of first-principle-based kinetic treatments, and global theory.  We emphasized  that,  although quite broad topics related to GAMs are investigated in the past two decades,    the    interest of the  fusion community on GAMs is due to their potential capabilities of regulating microscale turbulences and the associated anomalous transport. Consequently, the research  on GAMs is  carried out aiming toward the final goal of understanding the nonlinear dynamics of DWs  and transport in the presence of GAMs. 

In Sec. \ref{sec:GAM_linear},   an important concept of GAM is introduced, i.e., GAM continuous spectrum due to system nonuniformity, which leads to the generation of short scale mode structures and the breakdown of fluid description. As a result, kinetic treatment is required for the dispersion relation of short wavelength KGAM;  e.g., the Landau damping rate due to wave-particle resonance at short wavelength, and the  accurate prediction of the kinetic dispersiveness due to FLR and FOW effects; both playing important roles in the nonlinear interactions with DWs, as noted in Sec. \ref{sec:NL_GAM_DW}.

In Sec. \ref{sec:EGAM}, the resonant excitation and nonlinear saturation of EGAM are reviewed, using the analogy  to the well-known beam plasma instability (BPI). One crucial difference of the EGAM in three dimensional torus with respect to  BPI in a strongly magnetized   plasma is the EGAM radial mode structure due to the coupling to GAM continuous spectrum; leading to global mode structure and finite threshold condition.  Nonlinear interactions of EGAM and DWs are observed in numerical simulations, and thus, EGAM is considered as a potential active control for DW turbulences.  The  Dyson equation describing nonlinear saturation of EGAM due to wave-particle phase space nonlinearities is also derived, and  qualitative discussions of  the   phase space structure generation and secular nonlinear EGAM dynamics are made.

In Sec. \ref{sec:NL_GAM_DW}, the nonlinear excitation of GAM by DWs/DAWs is investigated, and it is shown by local theory that  short wavelength KGAM is preferentially excited. The theory based on $k_{\perp}\rho_{ti}\ll1$ expansion,  valid for GAM excitation by ITG DW, is then extended to $k_{\perp}\rho_{ti}\sim O(1)$ to discuss the  excitation by CTEMs and   by TAEs, where electro-magnetic nonlinearity associated with Maxwell stress is also considered. The global theory including kinetic dispersiveness of both DW and KGAM  and finite pump DW radial  scales shows that    the parametric instability, which is a convective amplification process on the short time scale,  becomes a quasi-exponentially growing absolute instability on the longer time scale,  when nonuniformity of DW drive, i.e., diamagnetic drift frequency, is taken into account. The qualitative change of the parametric process  further shows the importance of kinetic treatment and system nonuniformity in proper analysis of the DW nonlinear dynamics and the resultant transport level. 

In Sec. \ref{sec:self_NLty}, the nonlinear self-couplings of GAM/EGAMs are investigated; with GAM second harmonic generation as an additional channel for GAM dissipation, and  ZFZF generation as a channel for power transfer from GAM/EGAM to ZFZF. An important control parameter for  the nonlinear process is $k_r\rho_d/\epsilon$. Noting that both GAM and ZFZF can regulate DWs at different rates, nonlinear self-couplings of GAMs then have potential implications for  the nonlinear dynamics of DWs and thus, fluctuation induced transport.

Finally, in Sec. \ref{sec:framework}, a ``unified theoretical framework of GAM/EGAM" is constructed, consistently including of all the physics discussed through Sec. \ref{sec:GAM_linear} to \ref{sec:NL_GAM_DW}. It provides  outlooks for important and challenging problems related to GAM, including 1) nonlinear dynamics of the coupled GAM-DW system,  2) nonlinear dynamics of EGAM and 3) nonlinear interactions of EGAM and DW. These problems are at the cutting edge of fusion research and will be topics of interest for the next decade.

\newpage
\section*{Acknowledgments}
This work is dedicated to late academician Changxuan Yu.  This work is supported by        the National Science Foundation of China under grant Nos.  11575157  and 11235009,  the National Magnetic Confinement Fusion Research Program under Grants Nos. 2013GB104004 and 2013GB111004, Fundamental Research Fund for Chinese Central Universities under Grant No. 2017FZA3004,  EUROfusion Consortium under grant agreement No. 633053 and US DoE Grants.

\end{document}